\documentclass[a4paper,fleqn]{cas-dc}
\usepackage[numbers]{natbib}
\usepackage{graphicx}
\usepackage{amsmath,amssymb,amsthm,amsfonts}
\usepackage{url}
\usepackage{xurl}
\usepackage{algorithm}
\usepackage{bm}
\usepackage{booktabs}
\usepackage{multirow}
\usepackage{microtype}
\usepackage{array}
\usepackage{algorithmic}
\usepackage{enumitem}
\usepackage{mathtools}
\usepackage{makecell}
\usepackage{tikz}
\usetikzlibrary{arrows.meta,positioning,calc,decorations.pathreplacing,fit,backgrounds}
\theoremstyle{plain}
\newtheorem{theorem}{Theorem}
\newtheorem{proposition}[theorem]{Proposition}
\theoremstyle{definition}

\usepackage{tabularx}
\newcolumntype{Y}[1]{>{\raggedright\arraybackslash}p{#1}}
\usepackage{hyperref}
\ExplSyntaxOn
\cs_set:Npn \__first_footerline:
  { \group_begin: \small \sffamily \__short_authors: :~ { \rmfamily \itshape Preprint } \group_end: }
\ExplSyntaxOff

\newcommand{\E}{\mathbb{E}}
\newcommand{\Var}{\operatorname{Var}}
\newcommand{\Jh}{J^{\mathrm{h}}}
\newcommand{\Jb}{J^{\mathrm{b}}}
\newcommand{\Ev}{\mathcal{E}_\tau}
\newcommand{\Obs}{\mathcal{O}}
\newcommand{\Cand}{\mathcal{A}_\tau}
\newcommand{\pafter}{\pi_{\mathrm{after}}}
\newcommand{\dataset}{TE-GvA}
\newcommand{\sig}{$^{\ast}$}
\newcommand{\code}[1]{\texttt{#1}}
\newcommand{\repourl}{\url{https://github.com/ruzbahani/TE-GvA}}
\newcommand{\bibdoi}[1]{\href{https://doi.org/#1}{doi:\,\nolinkurl{#1}}}
\begin{document}

\shorttitle{Gather or Act? Belief-Level Counterfactual Labels for CPS Incident Response}
\shortauthors{Ali Mohammadi Ruzbahani}
\setcounter{page}{1}

\title[mode = title]{Gather or Act? Belief-Level Counterfactual Labels for Deadline-Aware Incident Response in Cyber-Physical Systems}

\author[1]{Ali {Mohammadi Ruzbahani}}[orcid=0009-0004-3287-552X]
\ead{ali.mohammadiruzbaha@ucalgary.ca}

\affiliation[1]{organization={Smart Cyber-Physical (SCPS) Lab at Department of Electrical and Software Engineering, University of Calgary},
    city={Calgary, AB},
    country={Canada}}

\begin{abstract}
When an industrial control system misbehaves, the defender often must choose between acting now on the most likely cause and first gathering evidence that separates two look-alike causes, while the process drifts toward a shutdown. Learning or evaluating such decisions requires the cost of every candidate action from the same situation, which no public cyber-physical security dataset provides. We present \dataset{}, a counterfactual decision dataset for fault-driven incident response, built on the revised Tennessee Eastman process with a simulated control network, expiring evidence and defensive tools. It holds 9,908 decision points from 1,990 simulated component faults and process disturbances, each with 4 to 35 candidate actions labelled by branched rollouts with common random numbers. Labels computed under the true hidden cause, which we call hindsight labels, make gathering evidence the best choice in only 1.1--1.3\% of decisions, because every rollout runs with the cause known. For the dominant look-alike pair, a failed network link versus a frozen I/O module, we therefore add pair-conditioned belief labels: twin episodes replay the defender's history under the other cause, and the outcomes are weighted by the exact posterior over the pair. With them, gathering is best in 15.5\% of ambiguous decisions in the natural distribution and in 33.1\% of ambiguous stress-set decisions without a backup channel. When the posterior is correct and the twin is a faithful counterpart, belief labels have the same regression target as hindsight labels and lower variance. In a pre-specified, single-look test evaluation, a gradient-boosted rule trained on belief labels has 11\% lower regret on the stress set than the same rule trained on hindsight labels (the primary comparison), and 27\% lower regret on ambiguous stress decisions without a backup channel, which remains significant after Holm correction over the 20 pre-specified model comparisons. With a backup channel, and in the natural distribution, the two label sets perform about the same. Soft labels tolerate moderate posterior errors, whereas hard labels that commit to one cause do worse than hindsight labels.
\end{abstract}

\begin{keywords}
cyber-physical systems\sep industrial control systems\sep fault-driven incident response\sep value of information\sep counterfactual labels\sep Tennessee Eastman process\sep benchmark dataset
\end{keywords}
\maketitle
\hypersetup{pdfauthor={Ali Mohammadi Ruzbahani},pdfsubject={},pdfkeywords={cyber-physical systems, incident response, counterfactual labels, Tennessee Eastman process}}

\section{Introduction}
\label{sec:introduction}
Suppose the operator of a chemical plant is told that the readings of one sensor group have stopped changing. Two causes fit this symptom. The network link that carries the group's measurements may have failed, in which case the controller keeps receiving the last values it was sent. Alternatively, the remote I/O module that digitizes the group may have frozen and keeps publishing the same values. The controller acts on the same held readings in both cases, so the process behaves identically. The repairs differ, however: the link takes about 40 minutes to repair, the module about an hour to replace, and the wrong repair leaves the plant running on stale measurements while it drifts toward a protective shutdown. Tracing the link takes two minutes and would settle the question. Should the defender trace first, or act on the more likely cause straight away?

This is a gather-or-act decision under a physical deadline, and incident response in cyber-physical systems (CPS) is full of them. Evidence does not stay valid for long: a window summary is useful for 4 to 15 minutes, a trace result for a quarter of an hour, and an alert for ten minutes unless the condition is reported again. Actions take time to start and to complete and compete for a shared executor. Once the plant trips, the cost is paid no matter how good the diagnosis was. CPS security and process-monitoring research has produced many detectors and several public testbed and simulation datasets~\cite{goh2017swat,ahmed2017wadi,shin2020hai,rieth2017}, and a body of work on response planning with explicit models~\cite{zonouz2014rre,miehling2018pomdp,hammar2022stopping}. Neither covers the case in between. A detector says whether something is wrong, and a planner needs a model of what each action will do. Learning a decision rule from data, or evaluating one, is best supported by data in which every candidate action has been tried from the same situation, so that its consequence can be compared with the alternatives without off-policy corrections. Incident logs record one action per situation, and the existing CPS datasets record none.

\textbf{Hindsight labels decide the question in advance.} In a simulator, the obvious way to label a decision point is to roll out every candidate from a snapshot and record its cost. We call these \emph{hindsight labels}, because every rollout runs in the world where the hidden cause is the one that actually occurred. For gather-or-act decisions this has a systematic consequence. Once the true cause is fixed, the right repair is almost always among the candidates and its label already shows that it works, so a gathering action can only add delay and tool cost, unless its result changes what the follow-up policy does in that same world. In our data, gathering beats the best direct action in 1.1\% (natural set) and 1.3\% (stress set) of decisions under hindsight labels, and in 29\% (natural) and 45\% (stress) of these the only difference is the cost of the tool itself. Scored with such labels, the best action is always one chosen with knowledge of the cause, so a benchmark cannot credit a well-timed investigation. A model trained on them targets the same conditional mean as one trained on the belief labels introduced below (Proposition~\ref{prop:rb}), but it learns from labels that are noisiest exactly where gathering matters. Hindsight optimization in planning has a similar blind spot: it scores actions on sampled futures in which the randomness is already resolved~\cite{yoon2008hindsight}, so nothing in it rewards finding out which future holds. Here the effect sits in the dataset.

\textbf{Belief-level labels.} What matters to the defender is the expected cost of an action given what it knows, $\E[J(a)\mid\Ev]$, where $\Ev$ is everything the defender has observed by decision time $\tau$, expired items included. For the link-versus-module ambiguity we estimate its \emph{pair-conditioned} version, $\E[J(a)\mid\Ev,\,C\in\mathcal{P}]$, where $\mathcal{P}$ is the pair of look-alike causes and the onset time is treated as known. Every affected episode gets a \emph{twin} episode that shares all random streams with it and replays the defender's actions, but has the other cause active. Each candidate is rolled out in both worlds with common random numbers, and the two outcomes are weighted by the posterior over the two causes. The posterior is computed analytically from the stochastic evidence the defender has seen (link-health reports, trace results and operator reports) and from deterministic evidence that rules one cause out. Under these \emph{pair-conditioned belief labels} (belief labels for short), gathering is the best choice in 15.5\% of the ambiguous decisions of the natural set, in 23.0\% of those of the stress set, and in 33.1\% of the stress set's ambiguous decisions in which no backup channel is available. Acting as if the cause were known costs 1.55 (natural) and 2.69 (stress) cost points per ambiguous decision on average when measured with belief labels, and it is the wrong choice in 29\% (natural) and 34\% (stress) of these decisions. Our contributions are as follows.

\textbf{Problem and dataset (Secs.~\ref{sec:problem} and~\ref{sec:dataset}).} We formalize deadline-aware gather-or-act decisions with expiring evidence and build \dataset{}, released with this article, a counterfactual decision dataset on the revised Tennessee Eastman process~\cite{downs1993,bathelt2015} with a simulated control network, telemetry and an executor for defensive tools. It has a natural distribution and a stress set that oversamples decisive cases, with 9,908 decision points, 143,447 labelled candidate actions and 926,764 simulated rollouts in total. The evidence tensors pass explicit leakage checks, and the dataset is documented with a datasheet~\cite{gebru2021datasheets}.

\textbf{Belief-level labelling (Secs.~\ref{sec:labels} and~\ref{sec:twins}).} We use twin episodes with an exact posterior over a pair of look-alike causes to give decisions pair-conditioned belief labels, and we show with a Rao--Blackwell argument that, when the posterior is correct and the label noise does not depend on which cause is true, the resulting labels have the same regression target as hindsight labels and a variance that is smaller by at least the between-cause term (Proposition~\ref{prop:rb}). Belief labels change the answer to the gather-or-act question in 15\% (natural) and 22\% (stress) of ambiguous decisions.

\textbf{Evaluation protocol (Sec.~\ref{sec:protocol}).} Our evaluation separates the gather-or-act choice from the choice within each action family, scores counterfactual family choices with each method's own ranking, uses cluster-bootstrap inference over episodes and reports every pre-specified comparison under both label sets. All test-split comparisons come from one pre-specified evaluation of the sealed test split, run with frozen networks whose validation predictions we first reproduced bit for bit and with gradient-boosted rules refitted deterministically on the training split. Exploratory analyses added afterwards use the training and validation splits only, and dataset statistics are computed on the development pool.

\textbf{Findings (Sec.~\ref{sec:results}).} On the held-out stress set, a decision rule trained on belief labels has 11\% lower regret than the same rule trained on hindsight labels (the pre-specified primary comparison), and 27\% lower regret on ambiguous decisions without a backup channel. This gain remains significant after Holm correction over the 20 pre-specified comparisons of E1--E3, whether decisions are scored with belief labels or with hindsight labels, which involve no posterior. With a backup channel the two label sets perform about the same. In the natural set, which has fewer high-stakes decisions, the gain is small and not significant. On the stress set the gain survives moderate errors in the posterior, whereas hard labels that commit to one cause are worse than hindsight labels (significantly so under belief scoring). A neural direct decision model that reads the raw evidence log fails to pick up the count-type and latest-item facts that rank actions against each other. A target-aligned evidence summary closes more than half of its gap, but on the stress set a gradient-boosted rule on the same summary remains significantly better than every network; in the natural set the learned rules cannot be told apart.

\textbf{Scope.} The incidents in \dataset{} are component faults and process disturbances, not attacks, and our claims are about fault-driven incident response. A responder faces the same decision at the start of a security incident: a held sensor group looks the same whether a link failed, a module froze or someone interfered with the readings, and the choice between collecting evidence and responding has to be made before the cause, malicious or not, is known. The labelling method does not depend on the cause being benign, but it does assume evidence generated by a known observation model that does not adapt to the defender; an adversary who forges or suppresses evidence would violate that assumption. Sec.~\ref{sec:discussion} discusses what carries over to security incidents and what does not.

\section{Related Work}
\label{sec:related}
\textbf{CPS security datasets.} Public CPS security datasets contain sensor and actuator time series from testbeds or simulators, together with labelled attack or fault intervals. SWaT and WADI come from water treatment and distribution testbeds~\cite{goh2017swat,ahmed2017wadi}, HAI from a hardware-in-the-loop control testbed~\cite{shin2020hai}, and Rieth et al.\ published a large set of Tennessee Eastman runs with labelled process faults~\cite{rieth2017}. These datasets support detection and diagnosis, but none records what a defensive action would have caused, so none can be used to learn or evaluate a response decision (Table~\ref{tab:datasets}). The Tennessee Eastman process~\cite{downs1993}, usually run under Ricker's decentralized control~\cite{ricker1996} and since 2015 also available in a revised implementation~\cite{bathelt2015}, is a common testbed for CPS attack studies~\cite{cardenas2011,krotofil2014stale}. Krotofil et al.\ showed that a denial-of-service attack that leaves the controller working with stale data can drive the process into an unsafe state if it is launched at the right moment, while a badly timed one has limited impact~\cite{krotofil2014stale}; the held-reading incidents of our ambiguous family rely on the same mechanism. Giraldo et al.\ survey physics-based attack detection~\cite{giraldo2018survey}, and Urbina et al.\ quantify the damage that stealthy attacks can still cause under detection~\cite{urbina2016}, which is the regime in which response decisions matter.

\begin{table*}[t]
\centering
\caption{Public datasets for CPS security and process monitoring compared with \dataset{}. $^{a}$Only for the link-versus-module confusion (pair-conditioned); hindsight labels elsewhere.}
\label{tab:datasets}
\small
\resizebox{\textwidth}{!}{%
\begin{tabular}{@{}lllcccc@{}}
\toprule
Dataset & Setting & Labels & \makecell{Counterfactual\\action outcomes} & Deadlines & \makecell{Expiring\\evidence} & \makecell{Belief-\\level labels}\\
\midrule
SWaT~\cite{goh2017swat} & water treatment testbed & attack intervals & -- & -- & -- & -- \\
WADI~\cite{ahmed2017wadi} & water distribution testbed & attack intervals & -- & -- & -- & -- \\
HAI~\cite{shin2020hai} & hardware-in-the-loop testbed & attack intervals & -- & -- & -- & -- \\
TE runs~\cite{rieth2017} & simulated chemical process & fault type and onset & -- & -- & -- & -- \\
\dataset{} (ours) & simulated process, network and tools & per-action cost, trip, timing & \checkmark & \checkmark & \checkmark & \checkmark$^{a}$ \\
\bottomrule
\end{tabular}}
\end{table*}

\textbf{Intrusion response and sequential defence.} Intrusion response systems choose countermeasures from alerts using rules, cost models or optimization~\cite{inayat2016irs}. Model-based work casts the response as a game~\cite{zonouz2014rre}, a partially observable Markov decision process (POMDP)~\cite{miehling2018pomdp} or an optimal stopping problem~\cite{hammar2022stopping}. In these formulations the defender's uncertainty is explicit in a belief state, and in the stopping formulation so is the choice between waiting for more evidence and acting, but they need a transition and observation model of the system, and their policies are optimized in that model. They are then evaluated in that model~\cite{miehling2018pomdp}, with a prototype built on an intrusion detection system~\cite{zonouz2014rre}, or in an emulation~\cite{hammar2022stopping}, and none of these settings provides the outcome of every candidate action from a common state. \dataset{} complements this line of work. It gives the consequences of actions in a high-fidelity simulator, against which the first decisions of model-based and learned policies can be compared from the same states for fault-driven incidents, and from which decision rules can be learned without a hand-built model.

\textbf{Value of information and active diagnosis.} The value of an observation is the reduction in expected decision cost that it buys~\cite{howard1966voi}. In control engineering, active fault detection designs inputs that make faults distinguishable while the plant is being controlled~\cite{simandl2009afd}. Our setting adds three things: evidence has a validity window, observation and repair compete for the time left before a deadline, and both are carried out by tools with random durations. We measure the value of information from rollouts in the true and the twin world, mixed with an analytic posterior over the ambiguous pair, rather than from a planning model of the plant.

\textbf{Hindsight, counterfactual rollouts and common random numbers.} Hindsight optimization solves sampled deterministic futures and averages the results~\cite{yoon2008hindsight}. Each sampled future reveals how the randomness resolves, so the averaged value is an upper bound on the optimal one~\cite{yoon2008hindsight} and nothing in it rewards finding out which future holds. Hindsight labels in a dataset share this property, and we show that it changes the answer in a large share of the decisions that matter. Common random numbers drive the compared alternatives with the same random inputs. They are a widely used way to reduce the variance of comparisons between simulated systems, although the reduction is guaranteed only under conditions such as monotonicity~\cite{glasserman1992crn}, so we measure it directly (Sec.~\ref{sec:rollouts}). We branch rollouts from a snapshot with common random numbers in the true world and in the twin world, so that both outcomes of an action are computed under the same noise. The belief label is a conditional expectation over the hidden cause, in the spirit of Rao--Blackwellization~\cite{blackwell1947}.

\textbf{Learning decision rules.} Regressing action costs and acting on their argmin is a predict-then-optimize approach, where regret matters more than prediction error~\cite{elmachtoub2022spo}. Our features are heterogeneous tabular features, a setting in which tree-based models still tend to beat deep networks on medium-sized data~\cite{grinsztajn2022tabular}. They are also partly missing, which the histogram-based gradient-boosted trees~\cite{friedman2001gbm} of scikit-learn~\cite{pedregosa2011sklearn} handle natively. Our direct model applies self-attention~\cite{vaswani2017attention} to sets of evidence items and is also evaluated as a seed ensemble. The evaluation follows several recommendations of Arp et al.\ for machine learning in security~\cite{arp2022dosdonts}: splits by scenario rather than by sample, explicit leakage controls, a sealed test set and simple reference baselines. In their terms it remains a lab-only evaluation (Sec.~\ref{sec:discussion}).

\section{Problem Formulation}
\label{sec:problem}
\textbf{Plant, incidents and evidence.} A plant runs under closed-loop control. At an unknown onset time, an incident with hidden cause $C$ begins (Sec.~\ref{sec:incidents} lists the causes). Without a correct response, some incidents drive the plant to a protective shutdown, which we call a \emph{trip}. The time at which this would happen without intervention is the incident's \emph{deadline}; the defender does not know it.

The defender observes an \emph{evidence log}. Each item $i$ has a source (window summary, alert, link-health report, tool result, operator report, receipt of the defender's own actions, and others), a payload, an event time, an arrival time $a_i$, a validity end $v_i$ and possibly a retraction time $r_i$. At decision time $\tau$ the \emph{valid set} is
\begin{equation}
\mathcal{V}_\tau=\{\,i : a_i\le\tau<\min(v_i,r_i)\,\}.
\end{equation}
Items that have expired, have been retracted or have not arrived yet are not evidence at $\tau$. The posterior over causes is determined by the defender's full history $\Ev$, which includes everything that arrived by $\tau$, expired items included. The decision models in this paper read only $\mathcal{V}_\tau$.

\textbf{Actions, executor and cost.} At $\tau$ the defender picks one action from a candidate set $\Cand$. An action is a tool (Table~\ref{tab:tools}) together with a target and arguments. Tools belong to four families: \emph{observe} ($\Obs$: trace a link, query the historian, verify a sensor), \emph{respond} (switch to a backup channel, restore the primary, repair a link, recalibrate a sensor, replace an I/O module, change production), \emph{monitor} (wait for a given time) and \emph{escalate}. An executor queues each action, checks its cited evidence and preconditions again when it starts, and issues \emph{receipts} (accepted, running, succeeded, rejected, and so on) that enter the evidence log. Tool durations are random.

The cost $J$ accrued over a horizon $H$ after $\tau$ adds up a lump for a trip, penalties for near-limit operation and lost production, the running cost of a backup channel and the effort of each tool (Table~\ref{tab:tools}). A trip ends the rollout, so its lump stands for all losses after the trip. Because the defender keeps acting after $\tau$, the cost of choosing $a$ at $\tau$ depends on what happens next. We fix this with a follow-up policy $\pafter$:
\begin{equation}
J(a\mid s_\tau)=\sum\nolimits_{t\in[\tau,\tau+H]}\mathrm{cost}_t\ \text{ under }\ (a,\pafter),
\end{equation}
that is, the cost accrued over $[\tau,\tau+H]$ when $a$ is taken first and $\pafter$ acts afterwards, where $s_\tau$ is the full simulator state. $J$ is random because of the plant, network, tool and operator noise after $\tau$.

\textbf{Decisions, regret and the value of gathering.} A decision rule maps the evidence to a candidate, $d(\mathcal{V}_\tau)\in\Cand$. Given labels $\hat J(a)$ for all candidates, its \emph{regret} at a decision point is
\begin{equation}
R(d)=\hat J\big(d(\mathcal{V}_\tau)\big)-\min_{a\in\Cand}\hat J(a).
\label{eq:regret}
\end{equation}
Regret is measured against the candidates that were tried and under $\pafter$; it is not a distance to a global optimum. The \emph{gather-or-act gap} is
\begin{equation}
\Delta=\min_{a\in\Obs\cap\Cand}\hat J(a)-\min_{a\in\Cand\setminus\Obs}\hat J(a),
\label{eq:gap}
\end{equation}
and \emph{gathering is optimal} when $\Delta<0$. We call the candidates in $\Cand\setminus\Obs$ \emph{direct actions}; besides repairs they include waiting and escalation. The observe family is defined by the tool, not by its informational effect: escalation also returns a diagnosis, and a switch to the backup channel or a repair reveals the cause through the data that follow. Sec.~\ref{sec:effects} measures how much this matters. Following Howard~\cite{howard1966voi}, we define the (net) value of information of an observation action $q$ as $\mathrm{VOI}(q)=\min_{a\notin\Obs}\hat J(a)-\hat J(q)$.

\subsection{Hindsight and Belief Labels}
\label{sec:labels}
Let $c^\star$ be the true cause. A \emph{hindsight label} estimates $\E[J(a)\mid s_\tau]$ by averaging $R$ rollouts from the snapshot at $\tau$, all in the true world:
\begin{equation}
\Jh(a)=\frac1R\sum_{k=1}^{R}J_k(a\mid c^\star).
\end{equation}
Now suppose that the defender's evidence cannot rule out a second cause $c'$ that produces the same observations. Let $\pi(c)=P(C=c\mid\Ev,\,C\in\mathcal{P})$ be the posterior restricted to the pair $\mathcal{P}=\{c^\star,c'\}$, and let $J_k(a\mid c',h)$ be the cost of $a$ in a \emph{twin world} in which $c'$ is active and the defender's history $h$ up to $\tau$ is unchanged. The \emph{belief label} is
\begin{equation}
\Jb(a)=\pi(c^\star)\,\Jh(a)+\pi(c')\,\frac1R\sum_{k=1}^{R}J_k(a\mid c',h).
\label{eq:belief}
\end{equation}
The two labels coincide outside such ambiguous decisions and, within them, whenever deterministic evidence has already excluded $c'$, in which case $\pi(c^\star)=1$.

Under the two assumptions of the following proposition, both label sets estimate the same regression target, and the belief labels do so with less noise.

\begin{proposition}[Belief labels as Rao--Blackwellized targets]
\label{prop:rb}
Fix a decision point and an action $a$. Let $\mathcal{I}$ be the information the labeller conditions on and $\pi(c)$ the posterior it computes for $c\in\{c_1,c_2\}$. Let $\hat J_c$ be the label estimate computed in the world with cause $c$ (the true world if $c=C$, the twin otherwise), with $\mu_c=\E[\hat J_c\mid\mathcal{I}]$ and $\sigma_c^2=\Var(\hat J_c\mid\mathcal{I})$. Assume that (A1) the posterior is correct, $\pi(c)=P(C=c\mid\mathcal{I})$, and (A2) the joint law of $(\hat J_{c_1},\hat J_{c_2})$ given $\mathcal{I}$ does not depend on which cause is true. Then, for $\Jh=\hat J_C$ and $\Jb=\sum_c\pi(c)\hat J_c$,
\begin{enumerate}[label=(\roman*),leftmargin=*]
\item $\E[\Jh\mid\mathcal{I}]=\E[\Jb\mid\mathcal{I}]=\sum_c\pi(c)\mu_c$;
\item $\Var(\Jb\mid\mathcal{I})\le\Var(\Jh\mid\mathcal{I})-\pi(c_1)\pi(c_2)(\mu_{c_1}-\mu_{c_2})^2$;
\item for every model input $X$ that is a function of $\mathcal{I}$, $\E[\Jh\mid X]=\E[\Jb\mid X]$. Squared-error regression on either label set therefore has the same population minimizer, and the belief labels remove at least $\E[\pi(c_1)\pi(c_2)(\mu_{c_1}-\mu_{c_2})^2\mid X]$ of label variance.
\end{enumerate}
\end{proposition}

The proof is in Appendix~\ref{app:proof}. The removed term is largest where gathering matters most. An action that is right under one cause and wrong under the other has a large $|\mu_{c_1}-\mu_{c_2}|$, often a full trip lump of 100 cost points, and its hindsight label reports one of the two outcomes at random. Proposition~\ref{prop:rb} is an application of the Rao--Blackwell argument~\cite{blackwell1947} and the law of total variance rather than a new statistical result: under (A1) and (A2), $\Jb=\E[\Jh\mid\mathcal{I},\hat J_{c_1},\hat J_{c_2}]$. What is new here is how we compute this conditional expectation, with twin episodes that replay the defender's history under the other cause, common random numbers and an exact posterior, and what it changes in the data and in the decision rules learned from them.

Proposition~\ref{prop:rb} also clarifies what each label set is good for. For \emph{training} with squared error, as in the GBM, both label sets have the same population target and the belief labels are less noisy; the direct model's Huber loss on $J$ is covered only approximately, because its minimizer depends on the whole label distribution and not only on its mean. For \emph{evaluation}, the per-decision quantities computed from hindsight labels (the best action, the gather-or-act gap and the value of information) are those of a defender who knows the cause, so the regret in Eq.~\eqref{eq:regret} under hindsight labels cannot recognize a well-timed investigation as optimal. The \emph{difference} in mean hindsight cost between two decision rules is a different matter. Each hindsight label is an unbiased estimate of $\E[J(a)\mid s_\tau]$, and each rule chooses from the evidence alone and is fitted on other data, so this difference is an unbiased estimate of the difference in expected cost $\E[J(d(\mathcal{V}_\tau))]$ at the evaluated decision points; it needs neither (A1) nor (A2). We therefore report every pre-specified comparison under both label sets. Belief labels are the primary view, and hindsight labels serve as a check that does not depend on the posterior or on the twin.

In \dataset{}, $\mathcal{I}$ consists of the defender's evidence $\Ev$ plus the fact that the incident is one of the ambiguous pair on a given sensor group. The labels are therefore \emph{pair-conditioned}. They average out the confusion inside the pair, but not the confusion between the pair and other causes, so they are belief-level labels for the confusion that dominates the decisive cases, not for every uncertainty the defender faces (Sec.~\ref{sec:discussion}). In practice the conditioning adds little: in the development pool, every decision at which a group-wide staleness signal is valid belongs to the pair on that group, and 86.4\% (natural) and 92.7\% (stress) of the ambiguous decisions have such a signal. In the others, including 36 natural decisions before the onset, the conditioning carries information the defender does not yet have. Assumption (A1) holds by construction, since the posterior uses the generator's own observation model, up to two approximations: the onset time is taken as known, and a tool result that completes in the same control step as an effect on the same link is scored against the state before that effect. Moving the assumed onset by one telemetry window (36\,s) changes the posterior by 0.006 to 0.018 on average and the belief-optimal action in 0.5--1.3\% of the ambiguous decisions not settled by deterministic evidence; averaging over an onset uniform within $\pm$3\,min changes the belief-optimal action in about 4\% (natural) and 3\% (stress). Sec.~\ref{sec:sensitivity} tests larger errors in the posterior.

Assumption (A2) concerns the state from which the twin's branches start. Let $\omega_e$ be the random draws that produce the cause-dependent stochastic evidence (for each telemetry window in which the link modes of the two worlds differ, whether the group's sensor link is reported and with what delay; the correctness of each trace result; the operator's answer), and let $\omega_s$ be all other draws. While the twin's cause has positive posterior probability, the deterministic evidence is the same function of $\omega_s$ and the replayed actions in both worlds, and $\omega_e$ is independent of $\omega_s$, so the shared $\omega_s$ is a valid draw for either cause given $\Ev$; the posterior rests on the same premise, which the labeller checks at every decision (Sec.~\ref{sec:twins}). The draws $\omega_e$ were conditioned on the true world's evidence, but after $\tau$ they enter only the twin's own trace, operator and link-health items, on which $\pafter$ never acts: its memory is the defender's history $h$, its rules ignore operator reports and the link-health reports of sensor links, and the executor checks only that cited items exist and are valid. All streams are reseeded at $\tau$, identically in both worlds. The twin world's own playbook would remember a different trace result in 20.0\% (natural) and 32.4\% (stress) of the ambiguous decisions, exactly those in which the defender holds a valid trace of the group's link taken while the link states of the two worlds differed. In these decisions $J(a\mid c',h)$, not the cost under the twin's own history, is the label that Eq.~\eqref{eq:belief} requires, and the posterior probability of the twin's cause is small (median 0.019 and 0.009). One further detail differs between the worlds: when the defender cited an item that exists only in the true world, such as a link-health report the twin never received, the twin's action cites evidence that both worlds share instead.

\section{The \dataset{} Dataset}
\label{sec:dataset}

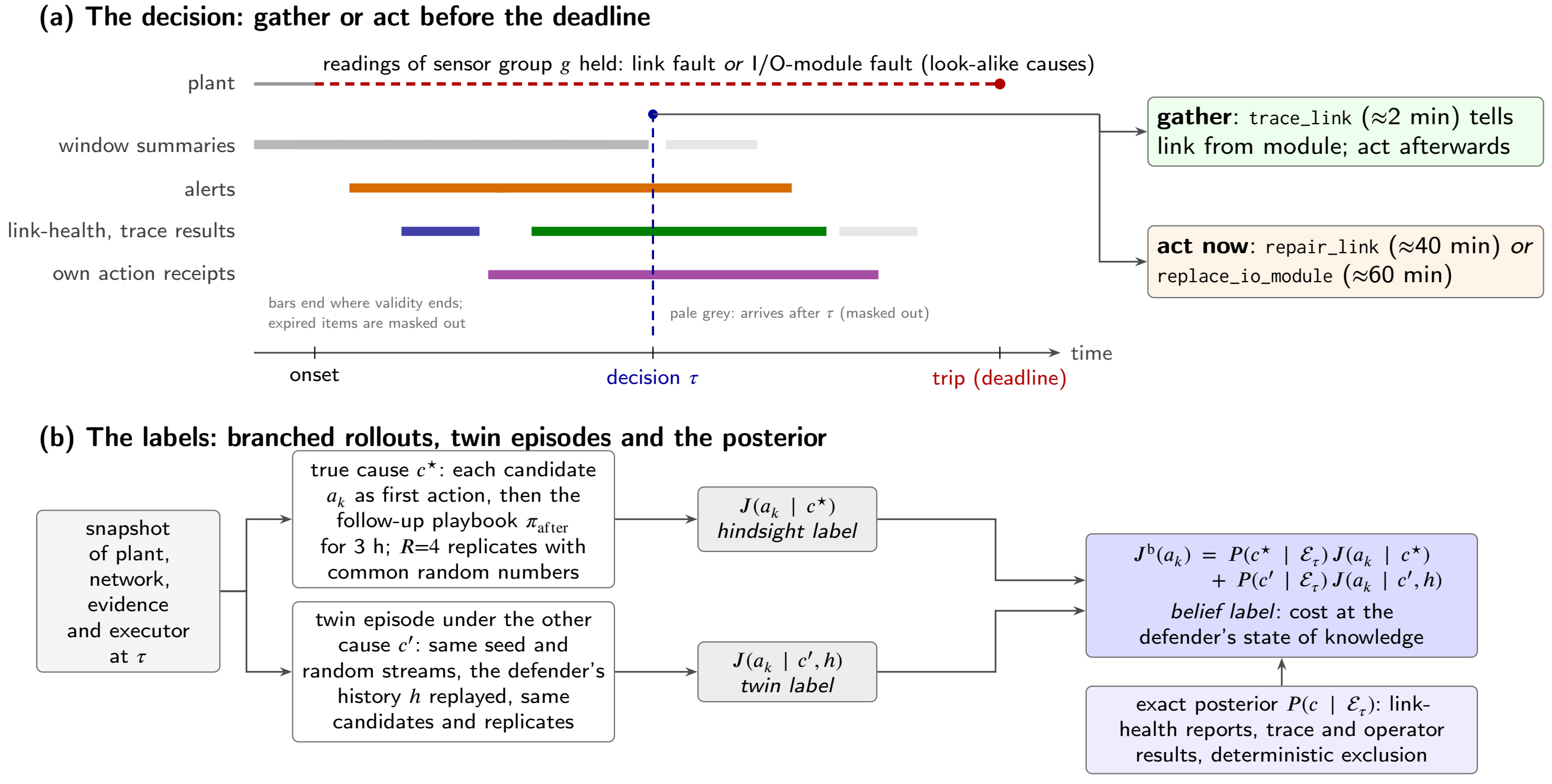
\begin{figure*}[t]
\centering
\resizebox{\textwidth}{!}{
\begin{tikzpicture}[
    font=\fontsize{8}{9.6}\selectfont,
    lane/.style={font=\fontsize{7}{8.4}\selectfont, anchor=east, text=black!75},
    item/.style={line width=3pt},
    box/.style={draw=black!55, rounded corners=2pt, align=left, inner sep=3pt},
    lab/.style={draw=black!55, rounded corners=2pt, align=center, inner sep=3pt, font=\fontsize{7}{8.4}\selectfont},
    arr/.style={-{Stealth[length=4pt]}, line width=0.55pt, black!70},
  ]
  \node[anchor=west, font=\fontsize{9}{10.8}\selectfont\bfseries] at (-2.6, 3.85) {(a) The decision: gather or act before the deadline};
  \draw[arr] (0,0) -- (9.3,0) node[right, font=\fontsize{7}{8.4}\selectfont] {time};
  \draw (0.7,0.07) -- (0.7,-0.07) node[below, font=\fontsize{7}{8.4}\selectfont] {onset};
  \draw (8.6,0.07) -- (8.6,-0.07) node[below, font=\fontsize{7}{8.4}\selectfont, text=red!70!black] {trip (deadline)};
  \node[lane] at (-0.1,3.1) {plant};
  \draw[line width=1.1pt, black!40] (0,3.1) -- (0.7,3.1);
  \draw[line width=1.1pt, red!70!black, densely dashed] (0.7,3.1) -- (8.6,3.1);
  \fill[red!70!black] (8.6,3.1) circle (1.8pt);
  \node[font=\fontsize{7}{8.4}\selectfont, anchor=south west, inner sep=1pt] at (0.75,3.15)
      {readings of sensor group $g$ held: link fault \emph{or} I/O-module fault (look-alike causes)};
  \node[lane] at (-0.1,2.4) {window summaries};
  \foreach \s in {0.0,0.5,1.0,1.5,2.0,2.5,3.0,3.5} {\draw[item, black!28] (\s,2.4) -- ++(1.05,0);}
  \draw[item, black!10] (4.75,2.4) -- ++(1.05,0);
  \node[lane] at (-0.1,1.9) {alerts};
  \draw[item, orange!85!black] (1.1,1.9) -- (2.9,1.9);
  \draw[item, orange!85!black] (2.8,1.9) -- (6.2,1.9);
  \node[lane] at (-0.1,1.4) {link-health, trace results};
  \draw[item, blue!55!black!75] (1.7,1.4) -- (2.6,1.4);
  \draw[item, green!50!black] (3.2,1.4) -- (6.6,1.4);
  \draw[item, black!10] (6.75,1.4) -- ++(0.9,0);
  \node[lane] at (-0.1,0.9) {own action receipts};
  \draw[item, violet!70] (2.7,0.9) -- (7.2,0.9);
  \draw[densely dashed, line width=0.7pt, blue!60!black] (4.6,0.2) -- (4.6,2.75);
  \fill[blue!60!black] (4.6,2.75) circle (1.6pt);
  \draw (4.6,0.07) -- (4.6,-0.07) node[below, font=\fontsize{7}{8.4}\selectfont, text=blue!60!black] {decision $\tau$};
  \node[font=\fontsize{5.5}{6.6}\selectfont, text=black!55, align=left, anchor=west] at (4.68,0.45) {pale grey: arrives after $\tau$ (masked out)};
  \node[font=\fontsize{5.5}{6.6}\selectfont, text=black!55, align=left, anchor=west] at (0.05,0.45) {bars end where validity ends;\\expired items are masked out};
  \node[box, anchor=west, fill=green!7, text width=4.35cm] (g) at (10.3,2.55)
      {\textbf{gather}: \texttt{trace\_link} ($\approx$2 min) tells link from module; act afterwards};
  \node[box, anchor=west, fill=orange!9, text width=4.35cm] (a) at (10.3,1.05)
      {\textbf{act now}: \texttt{repair\_link} ($\approx$40 min) \emph{or} \texttt{replace\_io\_module} ($\approx$60 min)};
  \draw[line width=0.55pt, black!70] (4.6,2.75) -- (9.75,2.75);
  \draw[arr] (9.75,2.75) |- (g.west);
  \draw[arr] (9.75,2.75) |- (a.west);
  \node[anchor=west, font=\fontsize{9}{10.8}\selectfont\bfseries] at (-2.6, -1.0) {(b) The labels: branched rollouts, twin episodes and the posterior};
  \node[lab, fill=black!4, text width=1.9cm] (snap) at (-1.45,-2.75) {snapshot of plant,\\network, evidence\\and executor at $\tau$};
  \node[lab, fill=white, text width=3.5cm] (prim) at (2.3,-1.92) {true cause $c^\star$: each candidate $a_k$ as first action, then the follow-up playbook $\pi_{\mathrm{after}}$ for 3 h; $R{=}4$ replicates with common random numbers};
  \node[lab, fill=white, text width=3.5cm] (twin) at (2.3,-3.68) {twin episode under the other cause $c'$: same seed and random streams, the defender's history $h$ replayed, same candidates and replicates};
  \node[lab, fill=black!7, text width=1.85cm] (Jh) at (6.15,-1.92) {$J(a_k\mid c^\star)$\\\emph{hindsight label}};
  \node[lab, fill=black!7, text width=1.85cm] (Jt) at (6.15,-3.68) {$J(a_k\mid c',h)$\\\emph{twin label}};
  \node[lab, fill=blue!14, text width=4.3cm] (Jb) at (11.85,-2.8) {$J^{\mathrm{b}}(a_k)=P(c^\star\mid\mathcal{E}_\tau)\,J(a_k\mid c^\star)$\\$\qquad\qquad+\;P(c'\mid\mathcal{E}_\tau)\,J(a_k\mid c',h)$\\[2pt]\emph{belief label}: cost at the defender's state of knowledge};
  \node[lab, fill=blue!7, text width=4.3cm] (post) at (11.85,-4.35) {exact posterior $P(c\mid\mathcal{E}_\tau)$: link-health reports, trace and operator results, deterministic exclusion};
  \draw[arr] (snap.east) -- ++(0.3,0) |- (prim.west);
  \draw[arr] (snap.east) -- ++(0.3,0) |- (twin.west);
  \draw[arr] (prim) -- (Jh);
  \draw[arr] (twin) -- (Jt);
  \draw[arr] (Jh.east) -- (8.55,-1.92) |- ([yshift=5pt]Jb.west);
  \draw[arr] (Jt.east) -- (8.55,-3.68) |- ([yshift=-5pt]Jb.west);
  \draw[arr] (post) -- (Jb);
\end{tikzpicture}}
\caption{(a) A gather-or-act decision: a failed link and a frozen I/O module produce the same held readings, and the decision rule sees only items valid at $\tau$. (b) Rollouts in the true and the twin world give the hindsight and the twin label; the belief label mixes them with the posterior given the defender's history $\Ev$.}
\label{fig:overview}
\end{figure*}

\dataset{} is produced by a pipeline that couples a process simulator, a control network with telemetry and monitoring, an executor for defensive tools, a behaviour policy that generates realistic decision points, and a labeller that branches the full simulator state (Fig.~\ref{fig:overview}).

\subsection{Simulator}
The process model is the revised Tennessee Eastman code of Bathelt, Ricker and Jelali~\cite{bathelt2015}, a modified version of the simulation program for the Downs and Vogel challenge problem~\cite{downs1993} that leaves the process model unchanged; the shutdown limits of the original model remain active. It runs in mode~1 under Ricker's decentralized multiloop controller~\cite{ricker1996} with a 1.8\,s control interval, 41 measurements and 12 manipulated variables (11 valves and the reactor agitator speed), and the kernel source is hash-checked every time the plant is constructed. Around the process we simulate the following components.
\begin{itemize}[leftmargin=*]
\item \emph{Control network.} Sensor and actuator links connect the field to the controller and are grouped by process unit (reactor, stripper, separator, feeds, recycle and analysers), with one remote I/O module per group. A failed link holds the last delivered values. A sensor link may have a backup channel, which delivers the same readings 36\,s late.
\item \emph{Telemetry and monitoring.} Every measurement is summarized over 36\,s windows whose summaries arrive with log-normal delays and stay valid for 4--15\,min. Link-health monitoring reports a down link with probability 0.05 per window and a healthy one with probability 0.0003. Staleness and deviation detectors are calibrated on normal operation only.
\item \emph{Executor.} Eleven tools (Table~\ref{tab:tools}) have log-normal durations and pass through a queue with 5\,s latency. Cited evidence and state freshness are checked again when a tool starts, and every step produces a receipt. Trace results are correct with probability 0.97. Escalation hands the incident to the operator, whose report names the true cause with probability 0.85, so escalation also returns discriminating evidence. The follow-up policy of the labeller (Sec.~\ref{sec:rollouts}) acts on trace results and on staleness that persists after a backup switch or a repair, but not on operator reports or field checks.
\item \emph{Policy layer.} Every repair must cite at least one item of world evidence, and the defender's own receipts never count. Repeated actions on the same target are rate-limited, and deep production cuts require a permission scope that the decision component does not have.
\end{itemize}

\begin{table}[t]
\centering
\caption{Tools, nominal durations and effort in cost points. Other costs: trip 100, near-limit operation 5 per hour, lost production up to 10 per hour, any sensor link on backup 0.5 per hour.}
\label{tab:tools}
\footnotesize
\setlength{\tabcolsep}{3.2pt}
\begin{tabular}{@{}llrr@{}}
\toprule
Tool & Family & Duration & Effort \\
\midrule
\code{monitor}$(T)$ (wait until $\tau+T$) & monitor & -- & 0 \\
\code{query\_historian} (recent readings) & observe & 10\,s & 0.04 \\
\code{trace\_link} (correct w.p.\ 0.97) & observe & 2\,min & 0.2 \\
\code{verify\_sensor} (field check of a group) & observe & 20\,min & 1 \\
\code{switch\_to\_backup} & respond & 1\,min & 0.2 \\
\code{restore\_primary} & respond & 1\,min & 0.2 \\
\code{repair\_link} & respond & 40\,min & 2 \\
\code{recalibrate\_sensor} (frozen, drift, bias) & respond & 60\,min & 2 \\
\code{replace\_io\_module} & respond & 60\,min & 2 \\
\code{set\_production} & respond & 30\,s & 0.2 \\
\code{escalate} (operator diagnosis) & escalate & 15\,min & 4 \\
\bottomrule
\end{tabular}
\end{table}

\subsection{Incidents and the Two Distributions}
\label{sec:incidents}
Each episode has a one-hour warm-up, an onset drawn uniformly between 10 and 60 minutes after it, and a six-hour horizon after the onset. Incidents are sensor and actuator link faults, I/O module faults, sensor faults (frozen, drift and bias) and process disturbances (the Tennessee Eastman IDVs), on top of a background disturbance (none, IDV8 or IDV1). There is no adversary. The two incident distributions share all code.
\begin{itemize}[leftmargin=*]
\item The \textbf{natural} set draws from the sampler's own distribution (link 35\%, I/O module 20\%, sensor 25\%, disturbance 20\%), and every sensor link has a backup channel. It is the headline distribution.
\item The \textbf{stress} set oversamples incidents in which both the trip time and the right repair depend on the evidence (link 35\%, I/O module 35\%, sensor 10\%, disturbance 20\%, with more weight on the reactor, stripper and feed units). Half of its episodes have no backup channel. In those episodes \code{switch\_to\_backup}, a cheap and reversible action that also exposes a frozen module, is not available, so a held sensor group becomes a real gather-or-act dilemma. The stress set is not a natural distribution, and we never pool it with the natural set.
\end{itemize}
An episode belongs to the \emph{ambiguous family} if its incident is a fault of a sensor group's link or of the same group's remote I/O module. We call the decisions of these episodes \emph{ambiguous decisions}; they make up 42.3\% of natural and 58.2\% of stress development decisions.

\subsection{Decision Points and Behaviour Policy}
The trajectories along which decisions are labelled come from a behaviour policy: a playbook with an \emph{act-first} and an \emph{investigate-first} mode, drawn per episode, plus uniform exploration over the candidate set with $\epsilon=0.3$ at each labelled point. The propensities of all submitted actions are logged. A decision point is labelled when a new alert arrives or, after the onset, when relevant evidence can be cited. The natural set has at most four labelled points per episode, of which at most one lies before the onset and serves as a false-alarm decision; the stress set has at most six, none before the onset. The candidate set is pruned from a registry of 109 actions to those whose target relates to the current evidence, plus the global actions (monitor, escalate and production changes). This leaves 13.8 (natural) and 14.9 (stress) candidates on average, and between 4 and 35 overall.

\subsection{Branched Rollouts with Common Random Numbers}
\label{sec:rollouts}
At every decision point the labeller takes a snapshot of the whole environment: the plant kernel with its noise generator, the controller, network, telemetry, evidence log, executor queue, cost accumulators and random streams. For each candidate and each of $R=4$ replicates, it restores the snapshot, reseeds every stream from (scenario seed, decision-point counter, replicate), submits the candidate as the first action, and lets $\pafter$ run for $H=3$\,h or until the episode ends, which happens earlier for 64 natural and 4 stress development decisions. Here $\pafter$ is a copy of the behaviour's playbook state switched to the act-first mode, and a first action \code{monitor}$(T)$ means that it waits strictly until $\tau+T$. The labeller records the cost and its components, the trip and its time, whether the cause was resolved, and the first action's final status, start time and duration, which is right-censored if the action is still running at the end of the branch. Replicate $k$ of every candidate uses the same noise path, so differences between candidates are estimated with common random numbers~\cite{glasserman1992crn}. The pooled variance of paired differences is 14.0\% (natural) and 22.1\% (stress) of what independent replicates would give. With $R=4$, each trip probability is a multiple of 0.25, and the runner-up is not separated from the best candidate by a paired 95\% interval over the replicates in 30.9\% (natural) and 34.5\% (stress) of the development decisions. Nearly all exact ties at the minimum under hindsight labels (41.5\% and 41.3\% of the decisions; 39.0\% and 37.5\% under belief labels) are between candidates whose costs agree in every replicate, mostly the two \code{monitor} durations (300\,s and 900\,s); they are ties of identical trajectories under common random numbers, not of noisy means. When the four replicates are resampled, the belief-optimal set of an ambiguous decision becomes disjoint from the original one in 12.7\% (natural) and 13.2\% (stress) of the resamples on average, and the gather-or-act answer flips in 5.7\% and 6.0\%; resampling four values understates their variance, so these shares are likely too low. This noise does not bias comparisons between decision rules: the labels of the scored decisions play no part in the rules' choices, and the episode bootstrap includes the noise.

\subsection{Twin Episodes and the Posterior}
\label{sec:twins}
Every episode in the ambiguous family is run a second time under the other cause of the pair. This twin shares the seed, and hence every random stream, as well as the background disturbance, the backup availability and the episode end. Two details make the twin a faithful counterpart.
\begin{itemize}[leftmargin=*]
\item \emph{Onset shift.} A link that fails at a control step holds the reading delivered at the previous step, whereas a module that freezes at a step holds the value it publishes at that step, which the controller receives one step later. We therefore start the module fault one control interval (1.8\,s) before the link fault. With this shift the delivered readings and all window summaries are bit-identical until the causes diverge; without it, 26\% of the window summaries differ in the example we measured.
\item \emph{Replayed history.} The twin is advanced to the exact control step of every submission in the primary episode and repeats it with the same tool, target and arguments. The twin is thus the world in which the same defender did the same things.
\end{itemize}
At every labelled $\tau$ the twin labels the same candidate list with the same replicates and reseeding. Its follow-up policy is a copy of the primary's $\pafter$ and so remembers what the defender actually observed, which gives $J(a\mid c',h)$ in Eq.~\eqref{eq:belief}.

We compute the posterior $\pi(c)=P(C=c\mid\Ev,\,C\in\mathcal{P})$ over the pair analytically. The prior is the sampler's probability of each incident, normalized within the pair:
\begin{equation}
P(\mathrm{link}\mid g)=\frac{q_L w_L(g)/W_L}{q_L w_L(g)/W_L+q_M w_M(g)/W_M},
\end{equation}
where $q_L$ and $q_M$ are the shares of link and I/O-module faults, $w_L(g)$ and $w_M(g)$ the sampler weights of group $g$'s sensor link and module, and $W_L$ and $W_M$ the total weights of all eleven links (sensor and actuator) and all six modules. Both sets give a group's sensor link and its module the same weight, so the prior does not depend on the group. It is $0.35\cdot8/(0.35\cdot8+0.20\cdot13)=0.519$ in the natural set and $0.35\cdot9.5/(0.35\cdot9.5+0.35\cdot15)=0.388$ in the stress set, where actuator links take 5 of the 13 and 5.5 of the 15 weight units of link faults. Stochastic evidence enters through its likelihood under each world's link state. This covers link-health windows in which the link modes of the two worlds differ (an observed report contributes its probability; a missing one contributes the probability that it was not issued or has not arrived yet, using the log-normal delay distribution), trace results and operator reports. Deterministic evidence (window summaries, alerts, historian queries, field checks and receipts) must be identical in both worlds. Once it differs, for instance because fresh data appeared after a repair, the twin's likelihood is zero and the belief label equals the hindsight label exactly. This happens in 15.3\% (natural) and 13.6\% (stress) of ambiguous decisions. Since the likelihood uses the generator's own parameters, the posterior is calibrated by construction; Fig.~\ref{fig:dataset}(c) confirms that the implementation behaves accordingly (expected calibration error, ECE, 0.015 and 0.014; 0.018 and 0.016 without the decisions in which deterministic evidence settled the posterior). In every subgroup we checked (backup availability, decision index, tercile of the time since the onset) the ECE lies between 0.009 and 0.027, below the 95th percentile expected from sampling noise under perfect calibration. Algorithm~\ref{alg:labels} summarizes the labelling of one decision point.

\begin{algorithm}[t]
\caption{Labelling one decision point of an ambiguous episode.}
\label{alg:labels}
\small\raggedright
\begin{algorithmic}[1]
\REQUIRE snapshot $S_\tau$ of the primary world (true cause $c^\star$); twin world $W'$ (cause $c'$) advanced to $\tau$ with the defender's history replayed; candidates $\Cand$; evidence $\Ev$
\FOR{$a\in\Cand$ and $k=1,\dots,R$}
  \STATE restore $S_\tau$, reseed all streams from (seed, decision, $k$), submit $a$, run $\pafter$ for $H$; $J_k(a\mid c^\star)\gets$ accrued cost
  \STATE restore $W'$ at $\tau$, reseed identically, submit $a$, run a copy of $\pafter$ (history $h$) for $H$; $J_k(a\mid c',h)\gets$ accrued cost
\ENDFOR
\IF{deterministic evidence in $\Ev$ differs between the worlds}
  \STATE $\pi(c^\star)\gets1$
\ELSE
  \STATE $\pi(c^\star)\gets$ prior $\times$ likelihood of the stochastic evidence, normalized over $\{c^\star,c'\}$
\ENDIF
\FOR{$a\in\Cand$}
  \STATE $\Jh(a)\gets\frac1R\sum_kJ_k(a\mid c^\star)$
  \STATE $\Jb(a)\gets\pi(c^\star)\Jh(a)+(1-\pi(c^\star))\frac1R\sum_kJ_k(a\mid c',h)$
\ENDFOR
\end{algorithmic}
\end{algorithm}

\subsection{Evidence Tensors and Leakage Controls}
Each decision point is stored as a fixed-size tensor of up to 256 evidence items (23 sources and 8 kinds) with their times and scaled payloads, filled round-robin over sources with the most recent items first. It contains every item valid at $\tau$ plus the items from the 30 minutes before $\tau$ and the two minutes after it. Expired and future items are included on purpose, so that a model ignoring the validity predicate would see future evidence: in the training split, 72.8\% (natural) and 71.6\% (stress) of the stored slots are invalid at $\tau$. Each candidate is described by 90 action features (tool, family, target, arguments, nominal duration, resource use and side-effect class).

The defender's own receipts count as evidence, and they are the main source of leakage. A receipt issued just after $\tau$ would reveal which action the behaviour policy chose, and a tool result in the look-ahead window would reveal its outcome. The builder therefore removes from each row every receipt and tool result of actions submitted at or after that decision, together with items derived from them. It also removes every item that is valid at $\tau$ according to its time stamps but was not yet known to the environment at $\tau$, such as a refusal stamped with its submission time but issued only after the queue latency. Automated checks over all development rows find no own-action receipts or tool results, no items of later actions and no receipts without an action identifier, and the arrays can be reproduced exactly from the raw episodes.

\subsection{Splits, Size and Statistics}
Splits are drawn by scenario. A fixed hash of the seed assigns each episode to train, validation or test (60/20/20), so all decisions of an episode and of its twin stay in the same split. Table~\ref{tab:stats} summarizes the dataset. The test split stayed sealed throughout development: no statistic, model choice or analysis looked at it until the single evaluation in Sec.~\ref{sec:results}. As in the datasheet, the descriptive statistics in this section are computed on the development pool (train and validation).

\begin{table}[t]
\centering
\caption{\dataset{} at a glance. Counts and wall time cover all splits, all other rows the development pool. ECE: expected calibration error; CRN variance ratio: variance of paired differences between candidates relative to independent replicates.}
\label{tab:stats}
\footnotesize
\setlength{\tabcolsep}{3pt}
\resizebox{\columnwidth}{!}{%
\begin{tabular}{@{}lrr@{}}
\toprule
 & natural & stress \\
\midrule
episodes (with labelled decisions) & 1,000 (990) & 1,000 (1,000) \\
decisions: train & 2,346 & 3,589 \\
\quad validation / test & 783 / 784 & 1,206 / 1,200 \\
labelled candidate actions & 54,324 & 89,123 \\
hindsight rollouts & 217,296 & 356,492 \\
twin rollouts & 114,320 & 238,656 \\
decisions in the ambiguous family & 42.3\% & 58.2\% \\
episodes without backup channel & 0\% & 51.9\% \\
high-stakes decisions (hindsight labels) & 9.7\% & 22.4\% \\
waiting is (co-)optimal (hindsight) & 66.5\% & 57.6\% \\
gathering optimal, hindsight, all & 1.1\% & 1.3\% \\
gathering optimal, belief, ambiguous & 15.5\% & 23.0\% \\
\quad with / without backup channel & 15.5\% / -- & 12.5\% / 33.1\% \\
twin excluded by deterministic evidence & 15.3\% & 13.6\% \\
posterior calibration: ECE / Brier & 0.015 / 0.125 & 0.014 / 0.092 \\
CRN variance ratio & 0.140 & 0.221 \\
generation wall time (15 workers) & 8.2\,h & 13.3\,h \\
\bottomrule
\end{tabular}}
\end{table}

\begin{figure*}[t]
\centering
\includegraphics[width=\textwidth]{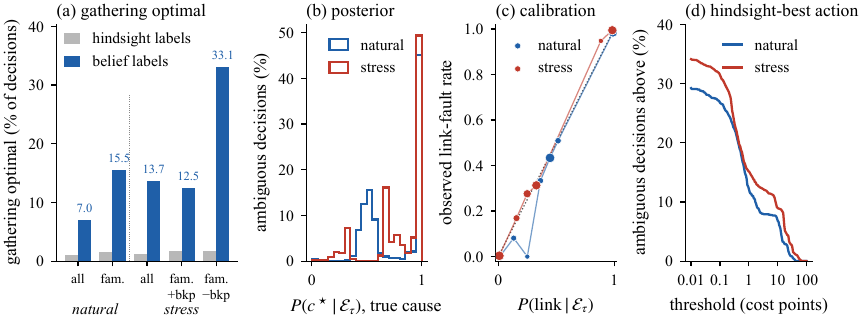}
\caption{Hindsight versus belief labels (development pool). (a) Share of decisions in which gathering is optimal; fam.: ambiguous family; $\pm$bkp: with or without a backup channel. (b) Posterior of the true cause in ambiguous decisions. (c) Calibration of $P(\mathrm{link}\mid\Ev)$ in bins with at least 10 decisions; larger markers: more decisions; dotted: perfect calibration. (d) Share of ambiguous decisions in which the hindsight-best action has a belief regret above the threshold.}
\label{fig:dataset}
\end{figure*}

Fig.~\ref{fig:dataset} shows what the belief labels change. Under hindsight labels, gathering is almost never optimal, with or without a backup channel. Under belief labels it is optimal in 15.5\% (natural) and 23.0\% (stress) of the ambiguous decisions, and without a backup channel the share nearly triples (from 12.5\% to 33.1\%), which is what the stress design was meant to produce. In 14.4\% and 21.4\% of ambiguous decisions, gathering is optimal under belief labels but not under hindsight labels; the reverse holds in only 0.45\% and 0.18\%. The action that hindsight labels declare best has a positive belief regret in 28.9\% (natural) and 34.2\% (stress) of ambiguous decisions, with 95th percentiles of 12.9 and 18.4 cost points.

\subsection{Which Actions Carry Information}
\label{sec:effects}
Table~\ref{tab:effects} groups the first actions of ambiguous decisions by their effect on the link-versus-module question rather than by tool family. Almost all of the gathering share comes from one action: tracing the suspect link accounts for 94\% (natural) and 97\% (stress) of the ambiguous decisions in which an observation is optimal under belief labels. Escalation and the field check of the suspect group are never optimal under either label set. The follow-up policy ignores their results, so their labels exceed that of a historian query on the same group by the difference in effort cost (median 3.96 and 0.96 cost points). Their information moves the posterior but has no value in the labels; a follow-up policy that acted on operator reports and field checks could make them worthwhile where no trace is available, which is the case in 9.6\% (natural) and 3.8\% (stress) of the ambiguous decisions. Counting the switch to the backup channel as information-producing raises the share in which such an action is strictly best, among the ambiguous decisions that offer one, to 26.7\% (natural) and 24.6\% (stress with a backup channel), but it raises the hindsight share as well (to 12.3\% and 13.7\%), because the switch helps under either cause. The difference between the two label sets therefore hardly changes, from 13.9 to 14.4 percentage points in the natural set and from 10.9 to 11.0 in the stress set with a backup channel (computed from unrounded shares). It shrinks to 1.8 and 0.9 percentage points only if committing to a repair is also counted as gathering information, because hindsight labels pick the right repair; without a backup channel 9.8 percentage points remain even then.

\begin{table}[t]
\centering
\caption{First actions in ambiguous development decisions by effect on the link-versus-module question (b: belief, h: hindsight labels). Upper part: share (\%) of decisions in which a row holds a co-optimal candidate (ties across rows count in each). Lower part: share (\%) of the decisions with candidates of the row in which its best candidate is strictly better than all others. Other observation: historian queries and checks of other targets; other response: all remaining responses.}
\label{tab:effects}
\footnotesize
\setlength{\tabcolsep}{3pt}
\begin{tabular}{@{}lrrrrrr@{}}
\toprule
 & \multicolumn{2}{c}{natural} & \multicolumn{2}{c}{\makecell{stress,\\backup}} & \multicolumn{2}{c}{\makecell{stress,\\no backup}} \\
\cmidrule(lr){2-3}\cmidrule(lr){4-5}\cmidrule(l){6-7}
 & b & h & b & h & b & h \\
\midrule
trace, field check, escalate & 14.5 & 1.3 & 11.6 & 1.5 & 32.4 & 1.3 \\
switch to backup & 11.9 & 11.1 & 13.0 & 12.2 & -- & -- \\
repair link or module & 11.9 & 24.6 & 18.6 & 28.7 & 25.8 & 53.1 \\
other observation & 1.0 & 0.5 & 0.9 & 0.3 & 0.6 & 0.5 \\
other response & 16.2 & 14.7 & 19.1 & 18.1 & 11.7 & 13.1 \\
wait (\code{monitor}) & 44.5 & 48.3 & 36.8 & 39.4 & 29.6 & 38.1 \\
\midrule
\multicolumn{7}{@{}l}{\emph{strictly best}} \\
\quad observe family & 15.5 & 1.6 & 12.5 & 1.7 & 33.1 & 1.8 \\
\quad first row or backup switch & 26.7 & 12.3 & 24.6 & 13.7 & 32.6 & 1.3 \\
\midrule
ambiguous decisions & \multicolumn{2}{c}{1,324} & \multicolumn{2}{c}{1,374} & \multicolumn{2}{c}{1,416} \\
\bottomrule
\end{tabular}
\end{table}

\section{Decision Rules}
\label{sec:models}
Every learned rule predicts a cost for each candidate and picks the argmin. None of them sees the hidden fields (cause, incident kind, onset, deadline, family membership or posterior).

\textbf{Fixed policies and references.} The two playbook modes and \emph{always wait} (monitor for 300\,s) are fixed policies, and a random candidate gives the scale. The \emph{oracle}, the argmin of the labels used for scoring, has zero regret by definition. The \emph{hindsight oracle} takes the argmin of the hindsight labels and is scored with belief labels; it knows the hidden cause and serves as a reference, not as a policy. A belief-threshold stopping rule, added after the test evaluation, is compared on the validation split only (Appendix~\ref{app:dev}).

\textbf{Gradient-boosted decision rule (GBM).} Each candidate is described by its 90 action features and by 151 hand-crafted features of the evidence valid at $\tau$, computed for the whole decision and at the candidate's own target. Per sensor group they include valid staleness and deviation alerts with their scores, the share of unchanged readings and field-check gaps; per link, health and loss reports and the latest trace result; further, the latest operator report, the number of valid items and the time since the first and the latest alert. Per tool, we count the valid receipts and note whether the latest one reports success. At the candidate's target we add the number of valid receipts, whether one of them is still running and the status of the latest receipt of the same tool. Of the 151 evidence features, 134 describe the decision as a whole and 17 the candidate's target. Missing evidence stays missing. A histogram gradient-boosted regressor~\cite{friedman2001gbm,pedregosa2011sklearn} with squared-error loss, 300 iterations, no early stopping and a fixed seed predicts $J$, optionally centred within each decision. Centring, the learning rate (0.05 or 0.1) and the number of leaves (15 or 63) are chosen by validation regret under the training labels. An \emph{action prior} applies the same boosting with fixed settings (centred, learning rate 0.05, 15 leaves) to the action features alone and amounts to an evidence-blind preference over actions.

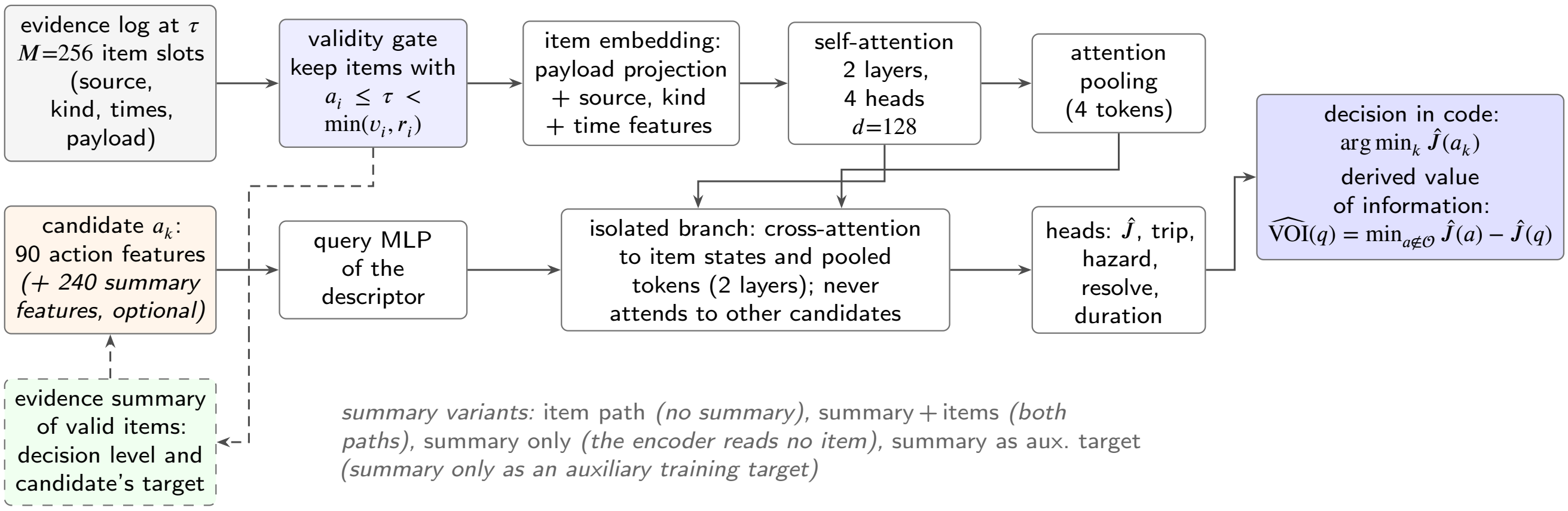
\begin{figure*}[t]
\centering
\resizebox{\textwidth}{!}{
\begin{tikzpicture}[
    font=\fontsize{7}{8.4}\selectfont,
    blk/.style={draw=black!55, rounded corners=2pt, align=center, inner sep=3pt, minimum height=0.95cm},
    arr/.style={-{Stealth[length=4pt]}, line width=0.55pt, black!70},
  ]
  \node[blk, fill=black!4, text width=2.0cm] (log) at (0,1.6) {evidence log at $\tau$\\$M{=}256$ item slots\\(source, kind, times,\\payload)};
  \node[blk, fill=blue!7, text width=1.75cm] (gate) at (2.75,1.6) {validity gate\\keep items with\\$a_i \le \tau < \min(v_i, r_i)$};
  \node[blk, fill=white, text width=2.05cm] (emb) at (5.45,1.6) {item embedding:\\payload projection\\+ source, kind\\+ time features};
  \node[blk, fill=white, text width=1.8cm] (set) at (8.1,1.6) {self-attention\\2 layers, 4 heads\\$d{=}128$};
  \node[blk, fill=white, text width=1.6cm] (pool) at (10.55,1.6) {attention\\pooling\\(4 tokens)};
  \draw[arr] (log) -- (gate);
  \draw[arr] (gate) -- (emb);
  \draw[arr] (emb) -- (set);
  \draw[arr] (set) -- (pool);
  \node[blk, fill=orange!8, text width=2.0cm] (cand) at (0,-0.35) {candidate $a_k$:\\90 action features\\\textit{(+ 240 summary\\features, optional)}};
  \node[blk, fill=white, text width=1.75cm] (q) at (2.75,-0.35) {query MLP\\of the\\descriptor};
  \node[blk, fill=white, text width=3.85cm] (br) at (6.75,-0.35) {isolated branch: cross-attention to item states and pooled tokens (2 layers); never attends to other candidates};
  \node[blk, fill=white, text width=1.6cm] (heads) at (10.55,-0.35) {heads: $\hat J$, trip,\\hazard, resolve,\\duration};
  \draw[arr] (cand) -- (q);
  \draw[arr] (q) -- (br);
  \draw[arr] (br) -- (heads);
  \draw[arr] (set.south) -- ++(0,-0.38) -| ([xshift=-0.6cm]br.north);
  \draw[arr] (pool.south) -- ++(0,-0.38) -| ([xshift=0.9cm]br.north);
  \node[blk, fill=blue!12, text width=3.0cm] (dec) at (13.6,0.62) {decision in code:\\$\arg\min_k \hat J(a_k)$\\[2pt]derived value of information:\\$\widehat{\mathrm{VOI}}(q)=\min_{a\notin\mathcal{O}}\hat J(a)-\hat J(q)$};
  \draw[arr] (heads.east) -- ++(0.3,0) |- (dec.west);
  \node[blk, fill=green!6, text width=2.0cm, dashed] (sum) at (0,-2.15) {evidence summary\\of valid items:\\decision level and\\candidate's target};
  \draw[densely dashed, line width=0.55pt, black!70] (gate.south) -- ++(0,-0.4) -| (1.45,-1.0);
  \draw[arr, densely dashed] (1.45,-1.0) |- (sum.east);
  \draw[arr, densely dashed] (sum.north) -- (cand.south);
  \node[font=\fontsize{7}{8.4}\selectfont\itshape, text=black!60, anchor=west, text width=8.5cm] at (2.3,-2.15)
      {summary variants: \emph{item path} (no summary), \emph{summary\,+\,items} (both paths), \emph{summary only} (the encoder reads no item), \emph{summary as aux.\ target} (summary only as an auxiliary training target)};
\end{tikzpicture}}
\caption{The direct decision model ($a_i$, $v_i$, $r_i$: arrival, validity end and retraction of item $i$). The dashed path adds the target-aligned evidence summary.}
\label{fig:model}
\end{figure*}

\textbf{Direct decision model.} The direct model (Fig.~\ref{fig:model}) reads the evidence once and answers one isolated query per candidate. A validity gate selects the items of $\mathcal{V}_\tau$ before any computation, so the output does not depend on the content of any invalid item, future items included. Valid items are embedded from their payload (one projection per source), from source and kind embeddings and from features of their times relative to $\tau$. Two self-attention blocks~\cite{vaswani2017attention} without positional encodings encode them, so the item states do not depend on the order of the items, and attention pooling with four learned queries compresses them into four tokens. Each candidate's descriptor is turned into a query that attends to the item states and pooled tokens in its own branch. Branches never attend to one another, so a candidate's prediction does not depend on which other candidates are on the menu. The heads predict $J$, the trip probability, a censoring-aware hazard of the trip time, resolution and the duration of the first action. With belief labels, only the $J$ and trip heads learn the belief labels; the hazard, resolution and duration heads keep the true-world outcomes. The decision is the argmin of the predicted $J$, computed in code, and the value of information of each observation is derived from the same predictions, which keeps the two consistent. Training uses a Huber loss on $J$ and a squared loss on $J$ centred within the decision, with early stopping on the latter on validation (Appendix~\ref{app:hyper}).

\textbf{Evidence summary variants.} During development we found that the learned item path does not recover the facts that rank actions against each other (Sec.~\ref{sec:repr}). We therefore evaluate four variants, each with five seeds: \emph{item path} (the model above); \emph{summary + items}, in which each candidate's descriptor also carries the 151 GBM evidence features (134 for the decision and 17 at the candidate's target), standardized with training statistics and clipped, plus 89 missingness indicators; \emph{summary only}, which is the same but with an encoder that reads no item; and \emph{summary as auxiliary target}, which has no summary in the input but an auxiliary head that predicts it from the branch state. Each variant is also scored as a five-seed ensemble that averages the predicted costs. Because the summary is computed from valid items only, the validity and menu-invariance guarantees still hold.

\section{Evaluation Protocol}
\label{sec:protocol}
\textbf{Metrics.} The primary metric is the regret of Eq.~\eqref{eq:regret} in cost points, averaged over decisions. We score each method under belief labels (primary) and hindsight labels (secondary), on all decisions and on subsets: the ambiguous family and, for the stress set, decisions with and without a backup channel. Regret splits into a \emph{family} part, which is the cost of the gather-or-act choice if the choice inside the chosen family were perfect, and a \emph{within-family} part. Because the family part assumes a perfect choice within the family, it misstates what a different gather-or-act choice would have cost a method that ranks actions poorly; we therefore also score counterfactual family choices with the method's own ranking inside the family. As descriptive gather-or-act diagnostics we report how often a method gathers; the \emph{costly-gather share}, the share of decisions with a labelled gap $\Delta\ge10$ (acting is at least 10 cost points better) in which the method still gathers; rank correlations between predicted and labelled gaps, which are not affected by the right-skewed mean of $\Delta$; and the \emph{act ranking}, the pooled Pearson correlation between predicted and labelled costs of the direct actions, each centred within its decision, over high-stakes decisions with at least three direct actions. High-stakes decisions are those in which the choice changes the trip probability under the scoring labels (belief labels unless stated otherwise; Table~\ref{tab:stats} uses hindsight labels), and for them we also report the regret and the trip probability of the chosen action.

\textbf{Statistics.} Decisions within an episode are correlated. All intervals are 95\% percentile bootstrap intervals that resample whole episodes, a pairs cluster bootstrap with episodes as clusters~\cite{cameron2008cluster} (2,000 replicates); with about 200 test episodes per set we are far from the regime of 5 to 30 clusters in which cluster-robust tests are known to over-reject. Methods are compared through paired differences on shared resamples, with a two-sided bootstrap $p=2\min\{P^\ast(d\le0),P^\ast(d\ge0)\}$ over the replicates $d$ of the difference (resolution 0.001). A seeded model is scored by its mean over seeds on each resample.

\textbf{Development and the frozen test evaluation.} Every modelling decision was made on the validation split. This includes the GBM grid, the direct model's architecture and training recipe (selected on an earlier version of the data and then kept), the diagnosis and the summary variants, which we confirmed on fresh seeds before opening the test split. We then wrote down and fixed the evaluation protocol: the methods, the pre-specified comparisons, the metrics and the commands. The comparisons are E1 (GBM trained on belief versus hindsight labels), E2 (evidence representations of the direct model), E3 (belief versus hindsight labels for the direct model) and the sensitivity grid of Sec.~\ref{sec:sensitivity}. Before predicting the test split, we reloaded the saved weights of all 52 networks (12 training runs: the four E2 variants with five seeds and the two E3 label sets with three seeds, each on both sets) and reproduced their saved validation predictions bit for bit (maximum absolute difference 0). The gradient-boosted rules and action priors are deterministic; they were refitted on the training split with the same code and grid, chose the same hyperparameters again and reproduced the development validation regrets (natural 1.128 and 1.219, stress 1.751 and 2.099, under belief labels). The test split was then evaluated once. Every pre-specified comparison is reported here, and every number the evaluation produced is released with the artefacts. Where space allows we show validation numbers next to the test numbers; they are optimistic, because model selection used them.

\textbf{Confirmatory and exploratory analyses.} The protocol fixed one primary comparison, E1 scored with belief labels on all decisions, tested separately in the natural and the stress set. The other pre-specified comparisons are secondary: E1 scored with hindsight labels on all decisions, E1 on the ambiguous family under both scorings and, in the stress set, on ambiguous decisions without a backup channel, as well as E2, E3 and the sensitivity grid. All other views (the remaining backup-channel subsets, the high-stakes views and the gather-or-act diagnostics) are descriptive. Intervals and $p$ values are unadjusted. The protocol did not fix a multiplicity correction; after the test evaluation we applied Holm's procedure~\cite{holm1979} to the recorded test $p$ values to control the family-wise error rate at 0.05 (Sec.~\ref{sec:e1}), without further access to the test split. All other analyses added after the test evaluation use the development data only and are exploratory: the label and replicate diagnostics in Secs.~\ref{sec:labels}, \ref{sec:rollouts} and~\ref{sec:twins}, the action classes in Sec.~\ref{sec:effects}, the backup-channel analysis in Sec.~\ref{sec:e1}, the margin rule in Sec.~\ref{sec:gather}, and the belief-threshold rule and trip-penalty analysis in Appendix~\ref{app:dev}.

\section{Results}
\label{sec:results}

\begin{figure*}[t]
\centering
\includegraphics[width=\textwidth]{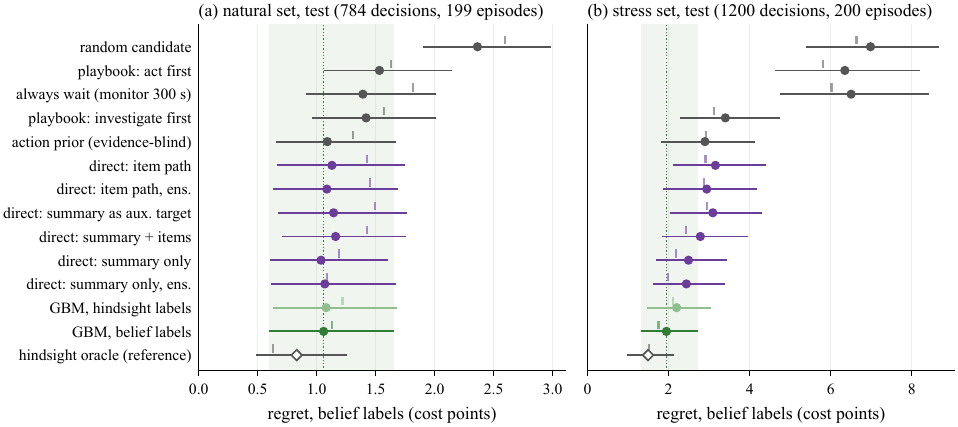}
\caption{Test regret under belief labels: point estimates with 95\% cluster-bootstrap intervals; ticks: validation estimates. Shaded band and dotted line: interval and estimate of the GBM trained on belief labels. Direct models: mean of five seeds; ens.: five-seed ensemble.}
\label{fig:results}
\end{figure*}

\begin{table*}[t]
\centering
\caption{Test results: regret (cost points) under belief ($R_\mathrm{b}$, with 95\% interval) and hindsight labels ($R_\mathrm{h}$), and belief regret and trip probability of the chosen action in high-stakes (HS) decisions. Direct models: mean of five seeds; ens.: five-seed ensemble. $^\ast$: differs from the GBM trained on belief labels (paired 95\% interval). Bold: best policy.}
\label{tab:main}
\small
\setlength{\tabcolsep}{4.5pt}
\begin{tabular}{@{}lccccccc@{}}
\toprule
 & \multicolumn{3}{c}{natural (784 decisions, 199 episodes)} & \multicolumn{4}{c}{stress (1,200 decisions, 200 episodes)} \\
\cmidrule(lr){2-4}\cmidrule(l){5-8}
Method & $R_\mathrm{b}$ & $R_\mathrm{h}$ & $R^\mathrm{HS}_\mathrm{b}$ & $R_\mathrm{b}$ & $R_\mathrm{h}$ & $R^\mathrm{HS}_\mathrm{b}$ & $p^\mathrm{HS}_\mathrm{trip}$ \\
\midrule
random candidate & 2.36\sig{} [1.90, 2.98] & 2.23 & 8.13 & 6.98\sig{} [5.38, 8.66] & 7.62 & 18.63 & 0.323 \\
playbook: act first & 1.53\sig{} [1.07, 2.15] & 1.35 & 6.30 & 6.35\sig{} [4.62, 8.20] & 7.05 & 18.04 & 0.328 \\
always wait (300\,s) & 1.39\sig{} [0.91, 2.01] & 1.14 & 7.60 & 6.50\sig{} [4.74, 8.41] & 7.20 & 19.37 & 0.342 \\
playbook: investigate first & 1.42\sig{} [0.96, 2.01] & 1.31 & 5.66 & 3.40\sig{} [2.29, 4.74] & 3.75 & 8.97 & 0.237 \\
action prior (evidence-blind) & 1.09 [0.66, 1.68] & 1.06 & 5.07 & 2.90\sig{} [1.81, 4.13] & 3.35 & 7.98 & 0.226 \\
\midrule
direct: item path & 1.13 [0.66, 1.75] & 1.06 & 5.62 & 3.16\sig{} [2.11, 4.40] & 3.60 & 8.87 & 0.235 \\
direct: item path, ens. & 1.09 [0.63, 1.69] & 1.01 & 5.38 & 2.94\sig{} [1.87, 4.19] & 3.33 & 8.17 & 0.227 \\
direct: summary as aux.\ target & 1.14 [0.67, 1.77] & 1.06 & 5.76 & 3.09\sig{} [2.03, 4.32] & 3.52 & 8.70 & 0.233 \\
direct: summary + items & 1.16 [0.71, 1.76] & 1.05 & 5.77 & 2.78\sig{} [1.83, 3.96] & 3.10 & 7.76 & 0.223 \\
direct: summary only & \textbf{1.04} [0.60, 1.60] & \textbf{0.96} & \textbf{4.96} & 2.49\sig{} [1.69, 3.44] & 2.89 & 6.82 & 0.213 \\
direct: summary only, ens. & 1.07 [0.61, 1.67] & 0.99 & 5.14 & 2.44\sig{} [1.62, 3.40] & 2.81 & 6.71 & 0.212 \\
\midrule
GBM, hindsight labels & 1.08 [0.63, 1.68] & 1.04 & 5.63 & 2.20\sig{} [1.48, 3.04] & 2.57 & 5.75 & 0.201 \\
GBM, belief labels & 1.06 [0.60, 1.66] & 0.97 & 5.44 & \textbf{1.95} [1.31, 2.72] & \textbf{2.33} & \textbf{4.89} & \textbf{0.192} \\
\midrule
hindsight oracle (reference) & 0.83 [0.48, 1.26] & 0 & 5.43 & 1.49 [0.98, 2.13] & 0 & 4.51 & 0.188 \\
\bottomrule
\end{tabular}
\end{table*}

\subsection{The Benchmark}
Fig.~\ref{fig:results} and Table~\ref{tab:main} give the test results. The two sets differ in difficulty by design. In the natural set, waiting is (co-)optimal in two thirds of the decisions, and outside the 15\% of test decisions that are high-stakes under belief labels (116 of 784; 8\% under hindsight labels) the costs within a decision lie close together. Every learned rule has lower regret than every fixed policy there; the GBM trained on belief labels improves on the investigate-first playbook by $-0.36$ [$-0.64$, $-0.11$], or 25\%. On 784 test decisions, however, the learned rules cannot be separated statistically from one another. In the stress set the playbooks lose between 3.4 and 6.3 cost points per decision, most of it in the 33\% of decisions that are high-stakes, and the methods separate clearly. The GBM trained on belief labels is the best learned rule. It reduces regret by 43\% compared with the investigate-first playbook ($-1.45$ [$-2.37$, $-0.70$]) and by 69\% compared with the act-first playbook. In high-stakes decisions it lowers the trip probability of the chosen action from 0.33 (act first) to 0.19, close to the 0.14 of the cost-optimal candidate. Measured with belief labels, it is statistically indistinguishable from the hindsight oracle ($+0.45$ [$-0.47$, $+1.39$]). The hindsight oracle itself has a positive belief regret of 0.83 and 1.49: knowing the cause and acting on it is not what belief labels reward, because in ambiguous decisions the belief-optimal action is often a different one, frequently an observation.

\subsection{Training on Belief Labels (E1, E3)}
\label{sec:e1}

\begin{figure*}[t]
\centering
\includegraphics[width=\textwidth]{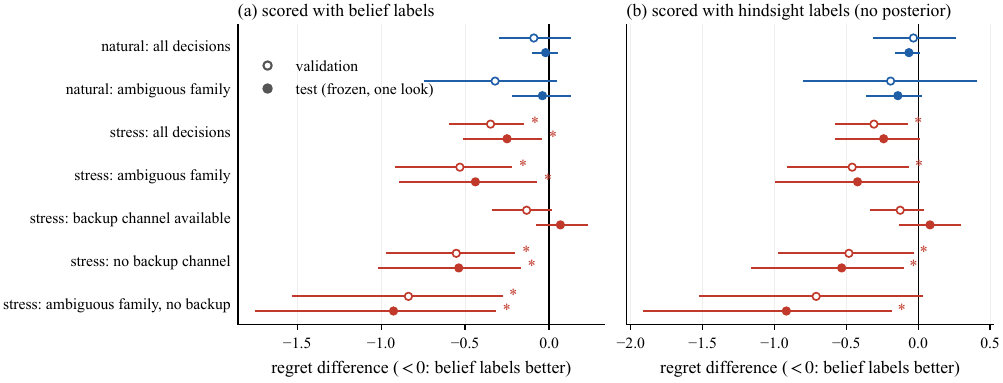}
\caption{E1: regret of the belief-trained minus the hindsight-trained GBM, with paired 95\% intervals. $^\ast$: the interval excludes zero (unadjusted).}
\label{fig:labels}
\end{figure*}

\begin{table*}[t]
\centering
\caption{E1 on the test split: regret of the belief-trained minus the hindsight-trained GBM (paired 95\% interval, bootstrap $p$, and in parentheses the change relative to the hindsight-trained GBM). Bold: the unadjusted interval excludes zero. Rows split by backup channel are descriptive, except the last.}
\label{tab:e1}
\small
\resizebox{\textwidth}{!}{%
\begin{tabular}{@{}llrcc@{}}
\toprule
Set & Decisions & $n$ & Scored with belief labels & Scored with hindsight labels \\
\midrule
natural & all & 784 & $-0.02$ [$-0.10$, $+0.05$], $p=0.63$ ($-2\%$) & $-0.06$ [$-0.16$, $+0.01$], $p=0.12$ ($-6\%$) \\
natural & ambiguous family & 336 & $-0.04$ [$-0.22$, $+0.13$], $p=0.65$ ($-2\%$) & $-0.14$ [$-0.36$, $+0.03$], $p=0.12$ ($-8\%$) \\
stress & all & 1,200 & $\bm{-0.25}$ [$-0.51$, $-0.04$], $p=0.016$ ($-11\%$) & $-0.24$ [$-0.58$, $+0.01$], $p=0.066$ ($-9\%$) \\
stress & ambiguous family & 726 & $\bm{-0.44}$ [$-0.89$, $-0.07$], $p=0.016$ ($-13\%$) & $-0.42$ [$-1.00$, $+0.01$], $p=0.058$ ($-11\%$) \\
stress & backup channel available & 570 & $+0.07$ [$-0.08$, $+0.23$], $p=0.41$ ($+3\%$) & $+0.08$ [$-0.14$, $+0.30$], $p=0.42$ ($+3\%$) \\
stress & ambiguous family, backup channel available & 348 & $+0.09$ [$-0.14$, $+0.37$], $p=0.49$ ($+3\%$) & $+0.12$ [$-0.23$, $+0.46$], $p=0.52$ ($+3\%$) \\
stress & no backup channel & 630 & $\bm{-0.54}$ [$-1.02$, $-0.17$], $p<0.001$ ($-24\%$) & $\bm{-0.53}$ [$-1.16$, $-0.10$], $p=0.003$ ($-19\%$) \\
stress & ambiguous family, no backup channel & 378 & $\bm{-0.93}$ [$-1.75$, $-0.32$], $p<0.001$ ($-27\%$) & $\bm{-0.92}$ [$-1.91$, $-0.18$], $p=0.002$ ($-21\%$) \\
\bottomrule
\end{tabular}}
\end{table*}

The training labels matter where the ambiguity matters (Fig.~\ref{fig:labels}, Table~\ref{tab:e1}). On the stress test set, the GBM trained on belief labels has 0.25 cost points (11\%) less regret than the same model trained on hindsight labels. Most of this comes from decisions without a backup channel ($-0.54$, 24\%), and within them from the ambiguous family ($-0.93$, 27\%), which is where Fig.~\ref{fig:dataset}(a) shows that gathering is most often right. Where a backup channel exists, the two label sets perform about the same. The test results reproduce the validation results in sign and size in every stress view except the two views with a backup channel, where all estimates are close to zero and change sign (all such decisions: $-0.13$ on validation and $+0.07$ on test; their ambiguous family: $-0.22$ and $+0.09$). Scored with hindsight labels, which involve neither the posterior nor the twin, the point estimates are nearly the same: $-0.24$ [$-0.58$, $+0.01$] ($p=0.066$) on all stress decisions, $-0.53$ [$-1.16$, $-0.10$] without a backup channel and $-0.92$ [$-1.91$, $-0.18$] in its ambiguous family. As argued after Proposition~\ref{prop:rb}, these paired differences are unbiased estimates of differences in expected cost. Under this posterior-free view the overall gain is not significant, but the gain where ambiguity peaks is. In the natural set all eight point estimates (two splits, two views and two label sets for scoring) favour belief labels, but none is significant. High-stakes decisions are fewer there (116 of 784 natural test decisions change the belief-level trip probability and 66 the hindsight-level one, against 391 and 277 of 1,200 in the stress set), and the noise of hindsight labels matters less when most decisions are clear.

For the direct model (E3: item path, three seeds, trained once on each label set), the training labels point the same way but matter less. On the test split under belief labels, belief minus hindsight labels is $-0.13$ [$-0.28$, $+0.02$] in the natural set and $-0.14$ [$-0.37$, $+0.07$] in the stress set; under hindsight scoring the natural-set difference is $-0.25$ [$-0.46$, $-0.07$]. Fifteen of the sixteen seed-mean point estimates (two sets, two splits and four views) favour belief labels; the exception is natural validation, ambiguous decisions under hindsight scoring ($+0.06$). The three-seed ensembles trained on the two label sets, however, are level: belief minus hindsight lies between $-0.07$ and $+0.18$ in the eight all-decision cells, and six of these are positive. In the stress set the effect is smaller than for the GBM ($-0.14$ against $-0.25$) and not significant, plausibly because the direct model's errors are dominated by its representation (Sec.~\ref{sec:repr}); in the natural set it is larger ($-0.13$ against $-0.02$).

With Holm's correction over the two primary tests, the stress-set gain remains significant ($p=0.016$ against a threshold of 0.025) and the natural-set gain does not. Over the sixteen views of Table~\ref{tab:e1}, the four views without a backup channel remain significant ($p\le0.003$), and over the 20 pre-specified comparisons of E1--E3 three do: E1 on ambiguous stress decisions without a backup channel under belief and hindsight scoring ($p<0.001$ and $p=0.002$), and the GBM's advantage over the summary-only ensemble in the stress set ($p<0.001$, Sec.~\ref{sec:repr}). The survivors with $p=0.002$ and $p=0.003$ lie closer to their Holm thresholds (0.0028 to 0.0038) than the Monte Carlo error of a 2,000-replicate bootstrap $p$ value. Adding the 60 cells of the sensitivity grid lowers the smallest Holm threshold to 0.000625, below the 0.001 resolution of the bootstrap $p$, so no comparison can then be certified.

\textbf{Backup channel and ambiguity (exploratory, development data).} A plausible reason why the two label sets perform alike where a backup channel exists is the follow-up policy, which switches to the backup channel after the first action and so limits what a wrong first action can cost; we did not measure how often this happens. The switch itself is co-optimal as a first action in 13.0\% of the ambiguous stress development decisions with a backup channel (Table~\ref{tab:effects}), but on the stress validation split the two GBMs almost never choose it first. They choose different actions in 30.1\% of the validation decisions with a backup channel and in 19.4\% of those without, yet a disagreement costs the hindsight-trained rule 0.44 cost points on average in the first case and 2.85 in the second. E1 is $-0.13$ with a backup channel and $-0.55$ without, a difference of $-0.42$ [$-0.84$, $-0.02$] (under hindsight scoring $-0.36$ [$-0.87$, $+0.15$]). Grouping the stress validation decisions by an estimate of the between-cause term of Proposition~\ref{prop:rb} from the realized labels, $\pi(c^\star)\pi(c')\big(\Jh(a)-\bar J(a\mid c',h)\big)^2$ averaged over the candidates $a$, where $\bar J(a\mid c',h)$ is the twin label of Eq.~\eqref{eq:belief}, gives E1 of $-0.09$ [$-0.27$, $+0.04$] where the term is zero and $-0.91$ [$-1.92$, $-0.13$] and $-0.77$ [$-1.47$, $-0.20$] in its upper two terciles (under hindsight scoring $-0.99$ [$-2.03$, $-0.20$] and $-0.44$ [$-1.49$, $+0.55$]). These bins select on the realized labels, so they illustrate the mechanism rather than test it. In the natural set no tercile of the term, on its own cut points, differs significantly from zero.

\subsection{Robustness to a Misspecified Posterior}
\label{sec:sensitivity}

\begin{figure*}[t]
\centering
\includegraphics[width=\textwidth]{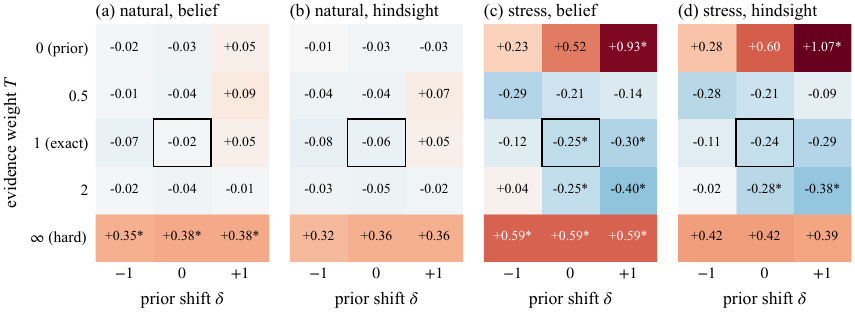}
\caption{E1 on all test decisions with a misspecified posterior: regret of the GBM trained on perturbed belief labels minus that of the hindsight-trained GBM, scored with belief (a, c) and hindsight labels (b, d). $T$: weight of the evidence; $\delta$: shift of the prior log-odds toward a link fault; $T=\infty$: hard label for the cause the evidence favours. Blue favours the perturbed labels; $^\ast$: the 95\% interval excludes zero; box: exact posterior.}
\label{fig:sensitivity}
\end{figure*}

Proposition~\ref{prop:rb} assumes a correct posterior, and \dataset{} computes it with the generator's own observation model. A practitioner who builds belief labels would have to estimate that model. To see how much this matters, we rebuilt the training labels with a misspecified posterior and refitted the GBM with its hyperparameters fixed to the nominal choice. This is possible because the twin label can be recovered exactly from the stored labels whenever $\pi(c^\star)<1$. Writing the log-odds of a link fault as $\ell=\ell_0+\Lambda$, with prior log-odds $\ell_0$ and evidence log-likelihood ratio $\Lambda$, the perturbed posterior is
\begin{equation}
\ell'=\ell_0+\delta+T\Lambda .
\end{equation}
$T$ under- or over-weights the evidence, and $\delta$ multiplies the prior odds by $e^{\delta}$, roughly a factor of 2.7 up or down. With $T=0$ all stochastic evidence is ignored. With $T\to\infty$ each decision gets a hard label for the cause that the evidence favours, the sign of $\Lambda$; the prior decides only when $\Lambda=0$. This is the maximum-likelihood cause rather than the most probable one, but in the training labels the two differ in only 10 of 2,346 natural and 1 of 3,589 stress decisions, so this row also stands for the hard labels of a pipeline that first diagnoses and then labels. Decisions in which deterministic evidence excluded the twin keep $\pi(c^\star)=1$. We fixed the grid and the code on validation before opening the test split.

Fig.~\ref{fig:sensitivity} shows the test results. Moderate misspecification does no measurable harm on all decisions. In the stress set, every cell with $T\in\{0.5,1,2\}$ is better than or level with hindsight labels; eight of the nine point estimates favour the perturbed belief labels, four are significant under belief scoring ($T\in\{1,2\}$, $\delta\in\{0,+1\}$), and two of them ($T=2$) are also significant under hindsight scoring. Two kinds of label do cause harm. \emph{Hard labels} that commit to one cause ($T=\infty$) are significantly worse than hindsight labels in both sets under belief scoring ($+0.59$ [$+0.18$, $+0.97$] stress, $+0.38$ [$+0.13$, $+0.68$] natural). Under hindsight scoring the point estimates are similar ($+0.42$ and $+0.36$), but the intervals include zero. Hard labels are as certain as hindsight labels without being correct, and they are wrong precisely in the ambiguous decisions. \emph{Prior-only labels} ($T=0$) are neutral in the natural set. In the stress set all three point estimates are positive ($+0.23$ to $+0.93$), and the cell with the prior shifted toward a link fault is significantly worse ($+0.93$ [$+0.05$, $+1.88$] at $\delta=+1$, and $+1.07$ [$+0.18$, $+2.05$] under hindsight scoring). In practice, belief labels should be soft and should weigh the evidence roughly correctly; an exact observation model is not required.

\subsection{What the Neural Model Needs (E2)}
\label{sec:repr}

\begin{figure*}[t]
\centering
\includegraphics[width=\textwidth]{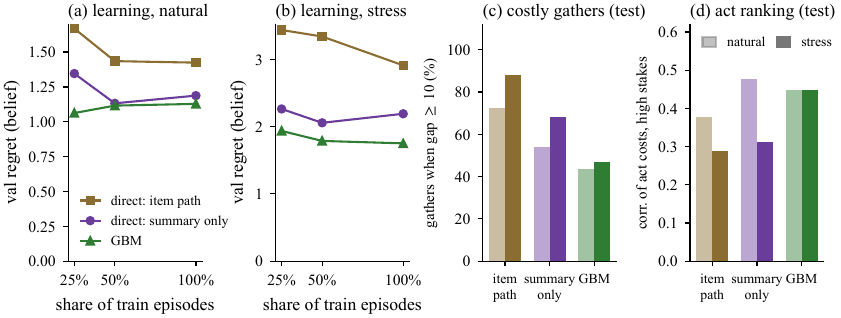}
\caption{(a, b) Validation regret against the share of training episodes. (c) Share of costly test decisions (acting is better by at least 10 cost points under belief labels; 16 natural, 75 stress) in which a method gathers. (d) Correlation of predicted and labelled direct-action costs, centred within each decision, in high-stakes test decisions. Direct models: mean over seeds; GBM: trained on belief labels. Light: natural; dark: stress.}
\label{fig:curves}
\end{figure*}

The direct model reads the raw evidence log and, unlike the GBM, needs no feature engineering. On the stress test set it is nevertheless the weakest learned rule (Table~\ref{tab:main}). With the learned item path alone, its regret of 3.16 is no better than that of the evidence-blind action prior (2.90) and 1.21 [0.49, 2.07] above the GBM. We diagnosed the cause on validation before opening the test split by testing four hypotheses, each re-derived by an independent second analysis.
\begin{itemize}[leftmargin=*]
\item The model does use the evidence on average. Removing all evidence turns it into the action prior (the natural gather share rises from 0.42 to 0.88), and its predicted gap ranks the decisions in which gathering is right (AUC 0.76).
\item It under-fits the cost differences between candidates rather than over-fitting. The bias is present in-sample on the training split from the second to fourth epoch onward, for every seed and both label sets, and predictions within a decision are shrunk: the slope of centred predicted on centred labelled cost is 0.25 (natural) and 0.19 (stress).
\item The mechanism is the act ranking. When the model cannot tell which repair is right, a cheap observation wins the argmin, so gathering acts as a hedge against a poor ranking of direct actions (Fig.~\ref{fig:curves}d). Changing the loss (squared error, a larger scale or weight of the centred term, centred loss only) shifted regret between the family and within-family parts but did not lower it, and a gather-or-act classification term removed the gather bias but collapsed the act ranking.
\item The bottleneck is the representation. A probe on the pooled state does not recover the family gap out of sample, while the facts that decide the act ranking, namely counts and the latest items at each candidate's target, are exactly what the GBM's summary contains.
\end{itemize}

The summary variants support this diagnosis on the stress test split (E2), although no E2 comparison among the networks survives Holm correction. Giving each candidate the target-aligned summary and dropping the item path lowers stress regret by 0.66 [0.03, 1.41] compared with the item path (seed means), and by 0.71 [0.06, 1.48] under hindsight scoring. As a five-seed ensemble it is 0.72 [0.02, 1.51] below the item path's seed mean, but only 0.50 [$-0.32$, 1.40] below the item path's own ensemble. It also reduces costly gathers from 88\% to 68\% (Fig.~\ref{fig:curves}c). Keeping the item path on top of the summary does not help ($+0.29$ [$-0.11$, $+0.82$] compared with summary only), and using the summary only as an auxiliary training target has no effect ($-0.06$ [$-0.21$, $+0.08$] compared with the item path). Telling the network which facts to extract is not enough; it has to be given them. In the natural set the variants cannot be distinguished. Even with the summary, the network does not reach the GBM on the stress set ($+0.49$ [$+0.16$, $+0.88$] for the ensemble), and its act ranking in high-stakes decisions stays weaker (0.31 against 0.45). The learning curves (Fig.~\ref{fig:curves}a,b) suggest that more data alone is not the remedy at the present scale: the GBM reaches its level with a quarter of the training episodes in the natural set and with half in the stress set, and the summary-only network stops improving at about half. The item path still improves on the stress set (from 3.34 to 2.91 between half and all of the data), so for it the curves cannot rule out gains from more data, but it starts far behind. What decides these choices is a small set of count-type and latest-item facts at each candidate's target; on medium-sized tabular data, tree-based models learn such irregular target functions and ignore uninformative features better than neural networks do~\cite{grinsztajn2022tabular}.

\subsection{Gather-or-Act Behaviour}
\label{sec:gather}
All learned rules gather more often than the belief labels prescribe; the belief-optimal share of test decisions is 7.9\% (natural) and 14.0\% (stress). Gathering too often costs little when the observation is cheap. What costs is gathering when acting is much better. In the stress test set the GBM gathers in 47\% of the 75 costly decisions and the item-path direct model in 88\%, and these decisions account for most of the difference in family regret between the two (0.75 against 2.14 cost points per decision). The GBM's predicted gap also ranks the labelled gap better (Spearman correlation 0.25 against 0.10). The accuracy of the binary gather-or-act choice is a poor summary: always acting gives the highest accuracy (0.86 stress and 0.92 natural) because gathering is the minority choice, yet the act-first playbook, which never gathers on the test split and reaches the same accuracy, has more than three times the GBM's regret in the stress set (6.35 against 1.95). Regret, its decomposition and the share of costly gathers are the informative measures.

For deployment, costly gathers matter in their own right. A conservative margin, under which the rule gathers only when the predicted cost of the best observation is at least $m$ points below that of the best direct action, costs little regret on the stress set (development data, Appendix~\ref{app:dev}). On the stress validation split, where the GBM gathers in 37\% of the costly decisions (47\% on test), $m=1$ lowers this share to 25\% and changes its regret by $+0.02$ [$-0.06$, $+0.10$]; cross-validated on the training split, regret is flat for $m\le1$, so the margin is a choice about risk rather than about average cost. In the natural set, where costly decisions are rare, a margin of 1 raises regret by $0.12$ [$0.05$, $0.22$].

\section{Discussion and Limitations}
\label{sec:discussion}
\textbf{Labels at the defender's level of knowledge.} For anyone building decision data for incident response, the main lesson is that labels must average over the causes that the defender cannot yet tell apart; otherwise the per-decision targets (the best action, the gather-or-act gap and the value of information) are those of a defender who knows the cause. Per decision, this bias does not average out, and it flips the gather-or-act answer in a fifth of the ambiguous stress decisions. As regression targets, hindsight labels are unbiased but carry the full between-cause variance (Proposition~\ref{prop:rb}), and at this data scale the policies learned from them are measurably worse where ambiguity is high, which is consistent with that variance but does not isolate it (Sec.~\ref{sec:e1}). Twin episodes are one way to compute such labels in a simulator, here for one pair of look-alike causes and conditional on that pair. They need a world model that can produce look-alike causes with shared randomness, which turned out to be manageable once the onset alignment was handled.

\textbf{Soft rather than hard labels.} The sensitivity analysis suggests a practical rule. Within the misspecifications we tested (evidence weight $T\in\{0.5,1,2\}$ and prior odds shifted by a factor $e^{\pm1}$), a soft mixture of per-cause outcomes was never significantly worse than hindsight labels on all decisions. Labelling each case with a single cause, the one its evidence favours, was significantly worse than ignoring the ambiguity altogether under belief scoring in both sets; under hindsight scoring the sign was the same but the intervals included zero. A diagnose-then-label pipeline, a natural engineering choice, falls into this second category.

\textbf{Engineered evidence features still win at this scale.} A neural model with structural guarantees (validity masking, menu invariance and a consistent value of information) is attractive for deployment, because its decisions can be checked against the evidence it used. On this benchmark it needs the same target-aligned evidence facts as the GBM, and even then it comes close to the GBM without matching it. The main bottleneck is extracting count-type and latest-item facts at each candidate's target from a log of expiring items. Even when those facts are supplied, the network maps them to cost less well than trees at this data size: on the stress test set the GBM's regret is 0.49 [0.16, 0.88] lower than that of the summary-only ensemble. Beyond the comparisons in this paper, \dataset{} supports survival modelling of action durations with censoring, evaluation of model-based response policies~\cite{miehling2018pomdp,hammar2022stopping} on fault-driven incidents against rollout ground truth, and the study of labelling methods.

\textbf{Relevance to security incident response.} The incidents in \dataset{} are faults, but the structure of the decision does not depend on intent. A held sensor group is also what a stale-data or denial-of-service attack on a control network produces~\cite{krotofil2014stale,cardenas2011}. The responder has to decide whether to investigate or to act before knowing whether the cause is malicious, and the constraints we model (expiring evidence, actions that take time and compete for an executor, a physical deadline) bind in the same way. Two findings should therefore carry over to security incidents. Labels computed under the true cause give almost no value to investigation whatever that cause is, so a benchmark of response decisions built from simulated attack scenarios needs labels that average over the causes the defender cannot yet distinguish as much as one built from faults. And a few count-type and latest-item facts about the evidence at each candidate's target decide the ranking of actions, which a learned rule has to be given or has to extract. What does not carry over is the observation model. An adversary who forges or suppresses evidence, or who adapts to the defender's actions, changes the likelihoods on which the posterior rests, and the twin construction would need a counterpart whose evidence is chosen rather than sampled. The robustness of soft labels to moderately misweighted evidence (Sec.~\ref{sec:sensitivity}) therefore carries over only where the evidence model is approximately known, which an adversary can break; this dataset does not cover that case. We therefore present \dataset{} as a benchmark for fault-driven incident response and the labelling method as a component that benchmarks with adversarial incident families can reuse.

\textbf{Cost weights and horizon.} A rollout stops at a trip, and neither the follow-up policy nor the posterior reads costs, so labels under any trip lump $L$ follow exactly from the stored ones as $J_L=J+(L-100)\,\hat p_{\mathrm{trip}}$, where $\hat p_{\mathrm{trip}}$ is the label's trip probability, for hindsight and belief labels alike. For $L$ between 25 and 400, gathering is optimal under belief labels in 14.8--15.9\% (natural) and 21.8--23.2\% (stress) of the ambiguous development decisions, and under hindsight labels in 1.1\% and 1.3\% of all development decisions at every $L$, while the belief regret of the hindsight-best action grows almost in proportion to $L$. With the hyperparameter grid rerun for each $L$, E1 on the stress validation split is negative for every $L\in\{50,100,200,400\}$ in all twelve views (six subsets, two scorings), and its interval on all decisions excludes zero at $L=100$ and $L=200$ (Appendix~\ref{app:dev}). The horizon cannot be varied from the stored labels, which keep each rollout's total cost and trip time but not its cost path. Only 24\% (natural) and 21\% (stress) of the trips within three hours occur in the first hour, so a one-hour horizon would remove most of the decisive cases; whether trips after three hours would change the labels requires new rollouts.

\textbf{Limitations.} All data are simulated. The process model is an established benchmark, but the network, telemetry timing, tools and costs are our design choices rather than a calibration of a real plant, and the cost weights are illustrative (the gather-or-act statistics are stable for trip lumps between 25 and 400, see above). Branched and twin rollouts cannot be collected on a real plant, so the dataset is a testbed for methods, not an estimate of real trip rates or deadlines. The incidents are faults and disturbances: evidence is never forged and nothing reacts to the defender.

Belief labels cover one confusion, link versus I/O module on a sensor group, conditioned on the pair. Decisions outside the pair (57.7\% of natural and 41.8\% of stress development decisions) carry hindsight labels, including other ambiguities such as a sensor fault versus a link, a disturbance versus a sensor fault, or whether an incident has started at all. The posterior is that of an idealized observer with the true observation parameters, a known onset and the conditioning on the pair (Sec.~\ref{sec:labels}). It conditions on everything the defender observed by $\tau$, including evidence older than the model's 30-minute context or beyond its 256 slots. The twin replays actions that were chosen for the primary's cause, so it trips far more often than the primary; it represents the cost of the defender's actual history under the other cause, as intended, not an independent sample of that cause. Costs are those of a first action followed by a fixed playbook for three hours, so the rollout labels cannot evaluate a planner's full multi-step behaviour, and regret is relative to a pruned candidate set. A different follow-up policy would change the labels, including the value of information, and a candidate that postpones a trip beyond the horizon looks as good as one that prevents it. In particular, evidence that the follow-up policy does not act on (operator reports, field checks and the link-health reports of sensor links) moves the posterior but has no value in the labels, which is why escalation and field checks are never optimal (Sec.~\ref{sec:effects}). With four replicates, the runner-up is not statistically separated from the best candidate in about a third of the decisions (Sec.~\ref{sec:rollouts}). Results of the historian and field-check tools are barely represented in the evidence, and labelled decisions concentrate in the first hour after the onset.

Our test-split baselines are fixed playbooks, an evidence-blind action prior and oracles. A belief-threshold stopping rule is compared on the validation split only (Appendix~\ref{app:dev}), because it was added after the test evaluation, and a particle-belief planner was not run. The natural set is the headline distribution, but its test split has only 116 decisions in which the choice changes the belief-level trip probability (66 under hindsight labels), too few to separate the learned rules, and the effect of the training labels there is small and not significant. The claims about training labels therefore rest on the stress set, where the effects without a backup channel replicate between validation and test. The primary comparison is one test per set, but the secondary comparisons and descriptive views are many. Sec.~\ref{sec:protocol} separates confirmatory from exploratory views, Sec.~\ref{sec:e1} gives Holm-adjusted results, and all computed views are released.

\section{Conclusion}
\label{sec:conclusion}
Deciding whether to gather evidence or act under a physical deadline is a central part of CPS incident response that has received little empirical study, largely because learning it requires data that security datasets do not contain: the consequences of every candidate action from the same situation. \dataset{} provides such data for fault-driven incidents on the Tennessee Eastman process. Building it exposed a pitfall that affects any simulator-labelled decision dataset whose rollouts run under the true hidden cause: such hindsight labels assume that the cause is known and therefore assign almost no value to finding it out. For the link-versus-module confusion, pair-conditioned belief labels from twin episodes remove this bias. When the posterior is correct and the twin is a faithful counterpart, they are provably lower-variance regression targets with the same minimizer. On the held-out stress set, where ambiguous high-stakes decisions are frequent, they lead to better decision rules (11\% lower regret overall, the pre-specified primary comparison, and 27\% lower on ambiguous decisions without a backup channel), while with a backup channel and in the natural set the two label sets perform about the same. The gain holds under moderate errors in the posterior but not under hard labelling. Future work includes adversarial incident families with their own observation models, belief labels for further confusions, and model-based planners evaluated against the rollout ground truth.

\section*{Code and Data Availability}
The code, the \dataset{} arrays with their metadata and datasheet, and the results behind every number in this article are available at \repourl{}; the data and the datasheet are archived on Zenodo~\cite{tegva2026zenodo}. The Tennessee Eastman kernel is not redistributed; it is available from its authors~\cite{bathelt2015}.

\section*{Declaration of Competing Interest}
The author declares no known competing financial interests or personal relationships that could have appeared to influence the work reported in this paper.

\section*{Acknowledgement}
While preparing this manuscript, I used AI-assisted tools to improve the fluency of the text and to correct grammar. I reviewed and edited the text afterwards, and I take full responsibility for the content of this article.

\appendix

\section{Proof of Proposition~\ref{prop:rb}}
\label{app:proof}
Write $\pi_c=\pi(c)$. In (i) and (ii), condition on $\mathcal{I}$ throughout. By (A1), $P(C=c\mid\mathcal{I})=\pi_c$, and by (A2), $(\hat J_{c_1},\hat J_{c_2})$ is independent of $C$ given $\mathcal{I}$.

(i) $\E[\Jh]=\E[\hat J_C]=\sum_c P(C=c)\,\E[\hat J_c\mid C=c]=\sum_c\pi_c\mu_c$, where the last step uses (A2). Since the weights $\pi_c$ are fixed given $\mathcal{I}$, $\E[\Jb]=\sum_c\pi_c\E[\hat J_c]=\sum_c\pi_c\mu_c$.

(ii) By the law of total variance and (A2),
\begin{equation*}
\begin{split}
\Var(\Jh)&=\E\big[\Var(\hat J_C\mid C)\big]+\Var\big(\E[\hat J_C\mid C]\big)\\
&=\textstyle\sum_c\pi_c\sigma_c^2+\pi_{c_1}\pi_{c_2}(\mu_{c_1}-\mu_{c_2})^2 .
\end{split}
\end{equation*}
For the belief label, with $\rho$ the correlation of $\hat J_{c_1}$ and $\hat J_{c_2}$,
\begin{equation*}
\begin{split}
\Var(\Jb)&=\pi_{c_1}^2\sigma_{c_1}^2+\pi_{c_2}^2\sigma_{c_2}^2+2\pi_{c_1}\pi_{c_2}\rho\,\sigma_{c_1}\sigma_{c_2}\\
&\le(\pi_{c_1}\sigma_{c_1}+\pi_{c_2}\sigma_{c_2})^2\le\pi_{c_1}\sigma_{c_1}^2+\pi_{c_2}\sigma_{c_2}^2 .
\end{split}
\end{equation*}
The first inequality uses $\rho\le1$, and the second is Jensen's inequality for the convex square with weights $\pi_{c_1}+\pi_{c_2}=1$. Combining the two displays gives (ii).

(iii) If $X$ is a function of $\mathcal{I}$, the tower property and (i) give $\E[\Jh\mid X]=\E[\E[\Jh\mid\mathcal{I}]\mid X]=\E[\E[\Jb\mid\mathcal{I}]\mid X]=\E[\Jb\mid X]$. The population minimizer of the squared error over functions of $X$ is the conditional mean, which is therefore the same for both label sets. The label noise around it is $\Var(\cdot\mid X)=\E[\Var(\cdot\mid\mathcal{I})\mid X]+\Var(\E[\cdot\mid\mathcal{I}]\mid X)$. The second term is the same for both label sets by (i), so the noise differs by $\E[\Var(\Jh\mid\mathcal{I})-\Var(\Jb\mid\mathcal{I})\mid X]$, which (ii) bounds from below. \hfill$\square$

The bound needs no assumption on the correlation between the two worlds; the hindsight label carries the between-cause term $\pi_{c_1}\pi_{c_2}(\mu_{c_1}-\mu_{c_2})^2$ in full.

\begin{table}[t]
\centering
\caption{Hyperparameters.}
\label{tab:hyper}
\footnotesize
\begin{tabularx}{\columnwidth}{@{}lX@{}}
\toprule
\multicolumn{2}{@{}l}{\emph{Direct decision model}}\\
width / heads & 128 / 4 \\
encoder layers / feed-forward & 2 / 256 \\
pooled tokens / branch layers & 4 / 2 \\
head width / dropout & 128 / 0.1 \\
parameters & 0.85\,M (item path), 0.89\,M (summary variants) \\
loss weights & $J$ (Huber, scale 10) 1, centred $J$ 1, trip 1, hazard 0.5, resolve 0.5, duration 0.2; auxiliary summary 0.3 (that variant only); ranking and gather-or-act terms 0 \\
optimizer & AdamW, learning rate $5\cdot10^{-4}$, weight decay $10^{-2}$, gradient clipping at 1.0 \\
batch / weight averaging & 32 decisions / exponential moving average, decay 0.995 \\
stopping & validation error of centred $J$, patience 15, at most 100 epochs \\
seeds & 3--7 (E2 variants and item path), 0--2 (E3) \\
training time & 3--13 min per seed on one NVIDIA RTX 5090 (several runs shared the GPU) \\
\midrule
\multicolumn{2}{@{}l}{\emph{Gradient-boosted rule}}\\
model & histogram gradient boosting regressor, squared-error loss, 300 iterations, at least 20 samples per leaf, L2 penalty 1.0 \\
grid (validation regret) & centring $\{$no, yes$\}$ $\times$ learning rate $\{0.05, 0.1\}$ $\times$ leaves $\{15, 63\}$ \\
chosen, belief labels & centred, 0.05, 63 leaves (natural); centred, 0.05, 15 leaves (stress) \\
chosen, hindsight labels & centred, 0.1, 63 leaves (natural); centred, 0.05, 15 leaves (stress) \\
action prior & action features only; centred, 0.05, 15 leaves (fixed, not tuned) \\
\bottomrule
\end{tabularx}
\end{table}

\begin{table}[t]
\centering
\caption{Validation-split regret under belief labels ($R_\mathrm{b}$) and hindsight labels ($R_\mathrm{h}$), as in Table~\ref{tab:main}.}
\label{tab:val}
\footnotesize
\setlength{\tabcolsep}{3pt}
\begin{tabular}{@{}lcccc@{}}
\toprule
 & \multicolumn{2}{c}{natural (783)} & \multicolumn{2}{c}{stress (1,206)} \\
\cmidrule(lr){2-3}\cmidrule(l){4-5}
Method & $R_\mathrm{b}$ & $R_\mathrm{h}$ & $R_\mathrm{b}$ & $R_\mathrm{h}$ \\
\midrule
playbook: act first & 1.63 & 1.94 & 5.82 & 6.29 \\
playbook: investigate first & 1.57 & 1.85 & 3.13 & 3.34 \\
action prior (evidence-blind) & 1.30 & 1.49 & 2.91 & 3.13 \\
direct: item path & 1.42 & 1.65 & 2.91 & 3.13 \\
direct: summary + items & 1.43 & 1.67 & 2.43 & 2.75 \\
direct: summary only & 1.19 & 1.42 & 2.19 & 2.42 \\
direct: summary only, ens. & 1.09 & 1.32 & 1.99 & 2.26 \\
direct: summary as aux.\ target & 1.49 & 1.76 & 2.95 & 3.17 \\
GBM, hindsight labels & 1.22 & 1.41 & 2.10 & 2.40 \\
GBM, belief labels & 1.13 & 1.38 & 1.75 & 2.09 \\
hindsight oracle (reference) & 0.63 & 0 & 1.53 & 0 \\
\bottomrule
\end{tabular}
\end{table}

\begin{table*}[t]
\centering
\caption{Belief-threshold rules on the validation split: regret under belief labels ($R_\mathrm{b}$) and paired difference to the GBM trained on belief labels, with 95\% interval. $^{a}$Uses the hidden exact posterior.}
\label{tab:bt}
\small
\begin{tabular}{@{}lcccc@{}}
\toprule
 & \multicolumn{2}{c}{natural} & \multicolumn{2}{c}{stress} \\
\cmidrule(lr){2-3}\cmidrule(l){4-5}
Method & $R_\mathrm{b}$ & difference & $R_\mathrm{b}$ & difference \\
\midrule
GBM, belief labels & 1.13 & -- & 1.75 & -- \\
threshold, exact posterior$^{a}$ & 1.46 & $+0.33$ [$-0.09$, $+0.88$] & 2.98 & $+1.23$ [$+0.46$, $+2.18$] \\
threshold, estimated posterior & 1.46 & $+0.33$ [$-0.09$, $+0.89$] & 3.05 & $+1.30$ [$+0.54$, $+2.25$] \\
gated GBM, exact posterior$^{a}$ & 1.16 & $+0.03$ [$-0.03$, $+0.10$] & 1.88 & $+0.13$ [$-0.04$, $+0.37$] \\
playbook: investigate first & 1.57 & $+0.44$ [$+0.00$, $+1.01$] & 3.13 & $+1.37$ [$+0.61$, $+2.32$] \\
\bottomrule
\end{tabular}
\end{table*}

\begin{table*}[t]
\centering
\caption{Sensitivity to the trip lump $L$ (development data). Gather: share (\%) of ambiguous decisions in which gathering is belief-optimal; $\bar R$: mean belief regret of the hindsight-best action in these decisions; E1: validation difference in belief regret (all decisions), with hyperparameters reselected for each $L$ (grid rerun) or kept at their $L=100$ choice (fixed); --: not computed. $^\ast$: the 95\% interval excludes zero.}
\label{tab:lump}
\small
\resizebox{\textwidth}{!}{%
\begin{tabular}{@{}rccccccccc@{}}
\toprule
 & \multicolumn{3}{c}{gather (\%)} & \multicolumn{2}{c}{$\bar R$} & \multicolumn{2}{c}{E1, natural} & \multicolumn{2}{c}{E1, stress} \\
\cmidrule(lr){2-4}\cmidrule(lr){5-6}\cmidrule(lr){7-8}\cmidrule(l){9-10}
$L$ & natural & stress & \makecell{stress,\\no backup} & natural & stress & grid rerun & fixed & grid rerun & fixed \\
\midrule
25 & 14.8 & 21.8 & 31.0 & 0.43 & 0.72 & -- & -- & -- & -- \\
50 & 15.4 & 22.7 & 32.7 & 0.80 & 1.37 & $-0.04$ [$-0.12$, $+0.05$] & $-0.05$ [$-0.13$, $+0.05$] & $-0.14$ [$-0.32$, $+0.02$] & $-0.10$ [$-0.27$, $+0.03$] \\
100 & 15.5 & 23.0 & 33.1 & 1.55 & 2.69 & $-0.09$ [$-0.30$, $+0.13$] & $-0.09$ [$-0.30$, $+0.13$] & $-0.35^\ast$ [$-0.60$, $-0.15$] & $-0.35^\ast$ [$-0.60$, $-0.15$] \\
200 & 15.7 & 23.2 & 33.4 & 3.05 & 5.34 & $-0.13$ [$-0.68$, $+0.41$] & $-0.08$ [$-0.38$, $+0.22$] & $-0.41^\ast$ [$-0.82$, $-0.05$] & $-0.59^\ast$ [$-1.05$, $-0.19$] \\
400 & 15.9 & 23.1 & 33.3 & 6.06 & 10.65 & $+0.45$ [$-0.30$, $+1.49$] & $-0.45$ [$-1.13$, $+0.17$] & $-0.39$ [$-1.15$, $+0.32$] & $-0.39$ [$-1.15$, $+0.32$] \\
\bottomrule
\end{tabular}}
\end{table*}

\section{Training Details and Validation Results}
\label{app:hyper}
Table~\ref{tab:hyper} lists the hyperparameters. The direct-model settings were selected on the validation split of an earlier version of the dataset (a grid over learning rate, dropout and width with three seeds) and kept unchanged. Table~\ref{tab:val} repeats the main comparison on the validation split, which was used for all modelling decisions; these numbers are optimistic for the learned rules.

\section{Development Analyses Added After the Test Evaluation}
\label{app:dev}

The analyses in this appendix use the training and validation splits only and are exploratory. The scripts and their outputs are released with the code.

\textbf{Belief-threshold rule.} The rule acts when a sensor group is held, that is, when at least two valid staleness alerts cover at least half of the group's channels (the playbook's criterion). It traces the group's link if $\max_c\pi(c)<\theta$ and no trace of that link is pending, waits while a trace is pending, and otherwise acts on the more likely cause (the act-first playbook's response to a link fault, or replacing the I/O module). In all other situations it follows the act-first playbook. The posterior is either the dataset's exact pair posterior, which makes the rule an idealized observer rather than a policy, or a logistic model of that posterior on the valid evidence at the group, fitted on the training split. The threshold $\theta$ is chosen by training regret, cross-fitted over five episode folds for the estimated posterior; it is 0.80 (natural) and 0.75 (stress) with the exact posterior and 0.75 and 0.95 with the estimated one. A gated variant uses the exact posterior only for the gather-or-act choice and lets the GBM choose within the chosen family. Table~\ref{tab:bt} gives the validation results. Each of the eight GBM grid settings has lower validation regret than either threshold rule. Of the two threshold rules, only the one with the exact posterior improves significantly on the investigate-first playbook ($-0.11$ [$-0.26$, $-0.003$] natural, $-0.14$ [$-0.27$, $-0.02$] stress; estimated posterior $-0.11$ [$-0.25$, $+0.002$] and $-0.07$ [$-0.17$, $+0.01$]). Most of the gap to the GBM comes from the rule's fixed choice of action rather than from its threshold, since the gated variant is level with the GBM.

\textbf{Margin rule.} The margin rule of Sec.~\ref{sec:gather} was evaluated with cross-fitted GBM predictions over six episode-grouped five-fold splits of the training set. Averaged over the splits, stress regret is 1.69--1.70 for margins between 0 and 1 and rises to 1.85 at $m=5$; the margin that a single split selects ranges from 0 to 1.5. On the validation split, margins of 0.25, 0.5 and 1 change stress regret by $-0.02$, $+0.02$ and $+0.02$ and lower the costly-gather share from 37\% to 36\%, 29\% and 25\%.

\textbf{Trip penalty.} Table~\ref{tab:lump} repeats the dataset statistics and E1 for other values of the trip lump $L$, computed exactly from the stored labels (Sec.~\ref{sec:discussion}). E1 is reported with the hyperparameter grid rerun on the validation split for each $L$ and with the hyperparameters fixed to their $L=100$ choices; the difference between the two shows how much selection on the same validation decisions adds to the spread at large $L$.

\bibliographystyle{IEEEtran}
\bibliography{references}

\end{document}